\documentclass{PoS}

\usepackage{amssymb, amsmath} 
\usepackage{epsfig}
\usepackage{graphicx}
\usepackage{cite}

 \newcommand{\GeV}{\mathrm{GeV}}

 \newcommand{\MS}{\overline{\sf MS}}

 \newcommand{\Athathat}{\hat{\hspace*{0mm}\hat{\tilde{A}}}}

 \newcommand{\Brack}[2]{\genfrac{[}{]}{0pt}{}{#1}{#2}}

\newcommand{\NN}{\nonumber}

\newcommand{\Li}{{\rm Li}}

\newcommand{\ep}{\varepsilon}

\title{{\footnotesize DESY 17-201, DO-TH 17/34}\\
The massive 3-loop operator matrix elements with two masses and the 
generalized variable flavor number 
scheme\thanks{This work was supported in part by the  Austrian Science Fund (FWF) grant SFB F50 (F5009-N15) and 
the European Commission through contract PITN-GA-2012-316704 ({HIGGSTOOLS}).}}

\ShortTitle{The massive 3-loop OMEs with two masses and the generalized VFNS}

\author{J.~Ablinger$^a$, J.~Bl\"umlein$^b$, \speaker{A.~De Freitas}$^b$, A.~Goedicke$^c$\thanks{A.~Hasselhuhn in previous publications.}, 
        C.~Schneider$^a$, K.~Sch\"onwald$^b$ and F.~Wi\ss{}brock$^{a,b}$\\ \\ \\
  \llap{$^a$}Research Institute for Symbolic Computation (RISC), Johannes Kepler University, \\ 
 Altenbergerstra\ss{}e 69, A--4040, Linz, Austria. \\
  \llap{$^b$}Deutsches Elektronen--Synchrotron, DESY, 
 Platanenallee 6, D-15738 Zeuthen, Germany. \\
  \llap{$^c$}Institut f\"ur Theoretische Teilchenphysik, Karlsruhe Institute of Technology (KIT), \\
 76128 Karlsruhe, Germany.}

\abstract{We report on our latest results in the calculation of the two--mass contributions to 3--loop operator 
matrix elements (OMEs). These OMEs are needed to compute the corresponding contributions to the deep-inealstic 
scattering structure functions and to generalize the variable flavor number scheme by including both charm and bottom 
quarks. We present the results for the non-singlet and $A_{gq,Q}$ OMEs, and compare the size of their contribution 
relative to the single mass case. Results for the gluonic OME $A_{gg,Q}$ are given in the physical case, going beyond 
those presented in a previous publication where scalar diagrams were computed. We also discuss our recently published 
two--mass contribution to the pure singlet OME, and present an alternative method of calculating the corresponding 
diagrams.}

\FullConference{RADCOR 2017 - 13th International Symposium on Radiative Corrections \\
  (Applications of Quantum Field Theory to Phenomenology), \\
		 September 24 - 29, 2017 \\
		 St. Gilgen, Austria}

\begin{document}

\allowdisplaybreaks

\section{Introduction}

\vspace*{1mm}
\noindent
Massive operator matrix elements (OMEs) constitute a key ingredient in the calculation of heavy quark corrections
to deep--inelastic scattering (DIS) structure functions at large virtualities, as they provide a link between the 
corresponding massless Wilson coefficients
and the massive ones. These OMEs are also needed to obtain the transition relations of the variable
flavor number scheme (VFNS).
Due to the accuracy of the currently available experimental data, the OMEs need to be calculated at three-loop order.
Their knowledge is of importance for the precise measurement of the strong coupling constant
$\alpha_s(M_Z^2)$~\cite{ALPHA},  of the parton distribution
functions \cite{PDF}, and of the heavy quark masses $m_c$ and $m_b$ \cite{MCMB} from the world deep inelastic data.
 
In a series of publications, we have computed the single mass contributions at three loops to the OME 
$A_{gq,Q}$~\cite{Ablinger:2014lka}, the $T_F^2$ terms of the gluonic OME $A_{gg,Q}$ \cite{Ablinger:2014uka}, the non-singlet 
contributions as well as all of the associated Wilson coefficients and structure functions 
\cite{Ablinger:2014vwa,Behring:2015zaa,Behring:2015roa,Behring:2016hpa}, the pure singlet result \cite{Ablinger:2014nga},
and the diagrams in the case of the ladder and $V$ topologies of the OME $A_{Qg}$ 
\cite{Ablinger:2015tua}. Furthermore, all diagrams contributing to $A_{Qg}$ which result from master integrals obeying 
first order factorizing differential equations have been completed \cite{Ablinger:2017ptf}, as well as all color and 
$\zeta$-value
terms, which can be determined using the method of arbitrarily large moments \cite{Blumlein:2017dxp}.
The logarithmic contributions to all OMEs were given in \cite{Behring:2014eya}. Before a series of moments has been
calculated for all massive OMEs in Ref.~\cite{Bierenbaum:2009mv}.

At three loops, irreducible Feynman diagrams with two fermion loops of different masses appear for the first 
time.\footnote{Reducible 2-mass contributions emerge already at NLO, cf.~\cite{VFNS}.}
The contributions from this type of diagrams cannot be ignored, since the mass of the bottom quark is not considerably 
larger than the mass of the charm quark, which in particular means that both quarks need to be decoupled simultaneously in the 
VFNS.
The renormalization of the OMEs in the 2-mass case has been performed in Ref.~\cite{Ablinger:2017err}. Here also 
the VNFS has been generalized to the 2-mass case. 

In these proceedings, we report on our recent progress in the calculation of these two-mass three-loop contributions
to the OMEs \cite{Ablinger:2017err, Ablinger:2017xml}. In Section~\ref{SSec-NS2MASS}, we study the simplest of these OMEs, 
namely,
$\tilde{A}_{qq,Q}^{(3), \rm NS}$ and $\tilde{A}_{gq,Q}^{(3)}$ (the tilde on top of the OMEs is used to denote the two--mass contributions). 
In these cases, the dependence of the OMEs on the Mellin variable $N$ and the masses fully factorizes. 
In Section~\ref{GGsection}, we show our recent results
on the physical diagrams for $\tilde{A}_{gg,Q}^{(3)}$, going beyond the results presented in \cite{Ablinger:2017err}, where a series of
scalar diagrams were computed. In Section~\ref{PSsection} we discuss the pure singlet case and show an alternative method 
of
computing the Feynman diagrams and the corresponding Feynman integrals to the one given in Ref.~\cite{Ablinger:2017xml} 
and conclude in Section~\ref{CONC}. 
\section{Operator matrix elements with a factorizing $N$ and $\eta$ dependence}{\label{SSec-NS2MASS}}

\vspace*{1mm}
\noindent
The simplest OMEs containing irreducible two--mass contributions are the three--loop non--singlet OME, $A_{qq,Q}^{(3), \rm 
NS}$, and the gluonic OME $A_{gq}^{(3)}$.
In the case of these two OMEs, the dependence on the ratio of the masses and the Mellin variable $N$ factorizes completely, 
unlike the case in all other OMEs, 
where these variables are intertwined in complicated functions, as we will see later.

All of the diagrams appearing in $A_{qq,Q}^{(3), \rm NS}$ and $A_{gq}^{(3)}$ contain two massive fermion bubbles, one of which may be rendered effectively massless by using a Mellin--Barnes
representation\cite{MB1a,MB1b,MB2,MB3,MB4}. 
\begin{eqnarray}
\raisebox{-6mm}{\includegraphics[keepaspectratio = true, scale = 0.8]{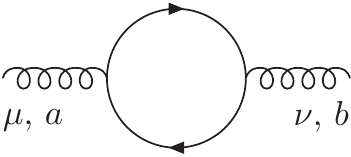}} &=& 
g_s^2 T_F \frac{4}{\pi} \left(4 \pi\right)^{-\ep/2} \left(k_{\mu}k_{\nu}-k^2 g_{\mu\nu} \right)
\nonumber \\ &&
\times
\int_{-i\,\infty}^{+i\,\infty} d \sigma
\left(\frac{m^2}{\mu^2}\right)^{\sigma} \left(-k^2\right)^{\ep/2 - \sigma}
\frac{\Gamma(\sigma-\ep/2) \Gamma^2(2-\sigma+\ep/2) \Gamma(-\sigma)}{\Gamma(4-2 \sigma+\ep)}.
\nonumber \\
\end{eqnarray}
This yields similar integrals as the ones appearing in the case
where there is one massive and one massless fermionic line \cite{Ablinger:2010ty,Blumlein:2012vq}. 
One may now combine the denominators of the Feynman integrals using Feynman parameters, integrate the momenta
and perform the Feynman parameter integrals in terms of Euler Beta--functions. 
The OMEs will then be given by a linear combination of contour integrals of the form 
\begin{eqnarray}
I &\propto& \Gamma \Brack{f_1(\ep,N),\ldots,f_i(\ep,N)}{f_{i+1}(\ep,N),\ldots,f_I(\ep,N)}
\NN \\ && \phantom{\times} \times
\int_{-i\,\infty}^{+i\,\infty} d\sigma \, \eta^{\sigma} \, 
\Gamma \Brack{g_1(\ep)+\sigma,g_2(\ep)+\sigma,g_3(\ep)+\sigma,g_4(\ep)
-\sigma,g_5(\ep)-\sigma}{g_6(\ep)+\sigma,g_7(\ep)-\sigma}, 
\label{NS2MASSContInt}
\end{eqnarray}
where the $f_j$ and the $g_j$ are linear functions, $N$ is the Mellin variable appearing in the operator insertion
Feynman rules, and $\eta$ is the ratio of the square of the masses
\begin{equation}
\eta = \frac{m_2^2}{m_1^2},
\end{equation}
where we assume $m_1 > m_2$, i.e., $\eta<1$.

After closing the contour in (\ref{NS2MASSContInt}) and collecting the residues,
the integrals end up being expressed as a linear combination of generalized hypergeometric $_4F_3$--functions \cite{HYP} 
\begin{eqnarray}
 I=\sum_j C_j\left(\ep,N\right){_4F_3}\Biggl[\genfrac{}{}{0pt}{}{a_{j,1}(\ep),a_{j,2}(\ep),a_{j,3}(\ep),a_{j,4}(\ep)}{b_{j,1}(\ep),b_{j,2}(\ep),b_{j,3}(\ep)},\eta\Biggr]~.
\label{I4F3}
\end{eqnarray}
Here and in the following finite and infinite sums occur which have to be summed. In the more involved cases
we thoroughly use difference field and ring theory algorithms \cite{Karr:81,
Schneider:01,
Schneider:05a,
Schneider:07d,
Schneider:10b,
Schneider:10c,
Schneider:15a,
Schneider:08c,
DFTheory} which are encoded in the package {\tt Sigma} \cite{SIG1,SIG2}.
Since in (\ref{I4F3}) the parameters of the hypergeometric functions depend only on the
dimensional regularization parameter $\ep$, their corresponding expansion may be
performed with the code {\tt HypExp 2} \cite{Huber:2007dx}. The results of these
expansions are then given in terms of the following (poly)logarithmic functions
\cite{POLYLOG1, POLYLOG2, Devoto:1983tc, Ablinger:2013jta}, 
\begin{equation}
\left\{\ln(\eta),~\ln(1-\eta),~\ln\left(\frac{1-\sqrt{\eta}}{1+\sqrt{\eta}}\right),~\Li_2\left({\sqrt{\eta}}\right),~
\Li_2\left(\eta\right),~\Li_3\left({\sqrt{\eta}}\right),~\Li_3\left(\eta\right)\right\}~.
\end{equation}

\subsection{The flavor non--singlet contribution}

\vspace*{1mm}
\noindent
The pole structure of the two--mass contribution to the non--singlet OME can be found from the renormalization procedure,
which involves mass, coupling constant and operator renormalization, as well as collinear factorization.
After the subtraction of the single-mass terms, one obtains \cite{Ablinger:2017err}
\begin{eqnarray}
 \Athathat_{qq,Q}^{(3), \rm{NS}} &=&
 -\frac{16}{3 \ep^3} \gamma_{qq}^{(0)} \beta_{0,Q}^2
 -\frac{4}{\ep^2}
\biggl[
\frac{2}{3} \beta_{0,Q} \hat{\gamma}_{qq}^{\rm{NS},(1)}
+\gamma_{qq}^{(0)} \beta_{0,Q}^2 \left(L_1+L_2\right)
\biggr] 
-\frac{2}{\ep} \biggl[
   \beta_{0,Q} \hat{\gamma}_{qq}^{\rm{NS},(1)} \left(L_2+L_1\right)
\NN\\&&
  +\gamma_{qq}^{(0)} \beta_{0,Q}^2 \left(L_1^2+ L_2 L_1+L_2^2\right)
  +4 a_{qq}^{\rm{NS},(2)} \beta_{0,Q}
  -\frac{1}{3} \hat{\tilde{\gamma}}_{qq}^{(2),\rm{NS}}
\biggr]
+\tilde{a}_{qq,Q}^{(3),\rm{NS}}~,
{\label{AthhNS3}}
\end{eqnarray}
where the $\gamma_{ij}^{(l)}$ are the anomalous dimensions at $l+1$ loops, $\beta_{0,Q} = -\frac{4}{3} T_F$, and
\begin{equation}
	L_1=\ln\left(\frac{m_1^2}{\mu^2}\right)~,\,\,\,L_2=\ln\left(\frac{m_2^2}{\mu^2}\right)~.
\end{equation}
The renormalized expression in the $\MS$--scheme, treating the heavy quarks in the on-shell scheme, is given by
\begin{eqnarray}
\tilde{A}_{qq,Q}^{(3), \MS, \rm{NS}} &=&
 \gamma_{qq}^{(0)} \beta_{0,Q}^2 \left(\frac{2}{3} L_1^3+\frac{2}{3} L_2^3+\frac{1}{2} L_2^2 L_1+ \frac{1}{2} L_1^2 L_2 \right)
+\beta_{0,Q} \hat{\gamma}_{qq}^{\rm{NS},(1)} \left(L_1^2+L_2^2\right)
\NN\\&&
+\left\{
4 a_{qq}^{\rm{NS},(2)} \beta_{0,Q}
+\frac{1}{2} \beta_{0,Q}^2 \gamma_{qq}^{(0)} \zeta_2
\right\} \left(L_1+L_2\right)
+8 \overline{a}_{qq}^{\rm{NS},(2)} \beta_{0,Q}
+\tilde{a}_{qq,Q}^{(3),\rm{NS}}~.
\end{eqnarray}
Both $\Athathat_{qq,Q}^{(3), \rm{NS}}$ and $\tilde{A}_{qq,Q}^{(3), \MS, \rm{NS}}$ vanish for $N=1$ at all orders in $\ep$ due to fermion number conservation.

\begin{figure}[h]
\begin{center}
\begin{minipage}[c]{0.20\linewidth}
  \includegraphics[width=1\textwidth]{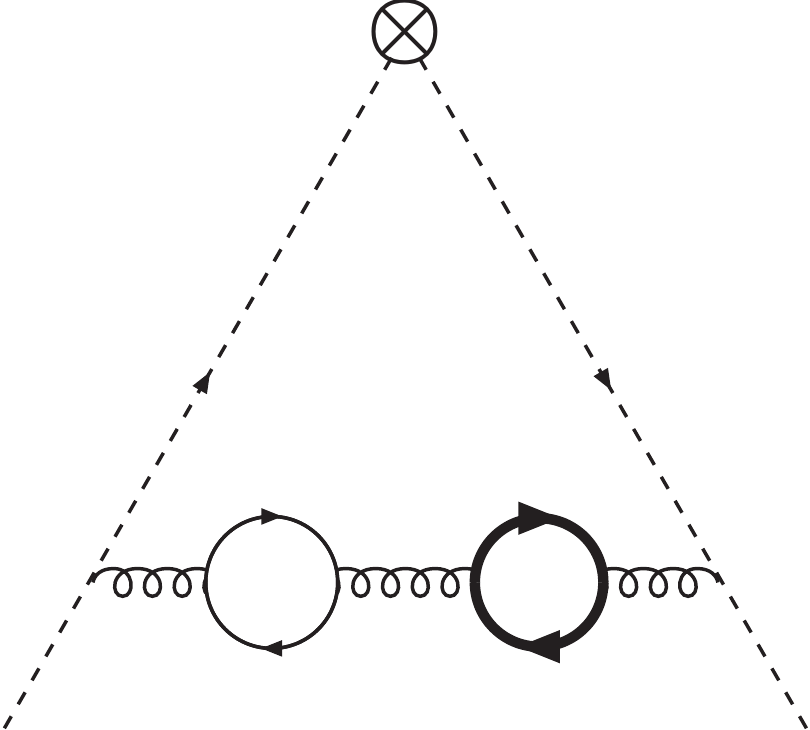}
\vspace*{-9mm}
\begin{center}
{\footnotesize (1)}
\end{center}
\end{minipage}
\hspace*{8mm}
\begin{minipage}[c]{0.20\linewidth}
  \includegraphics[width=1\textwidth]{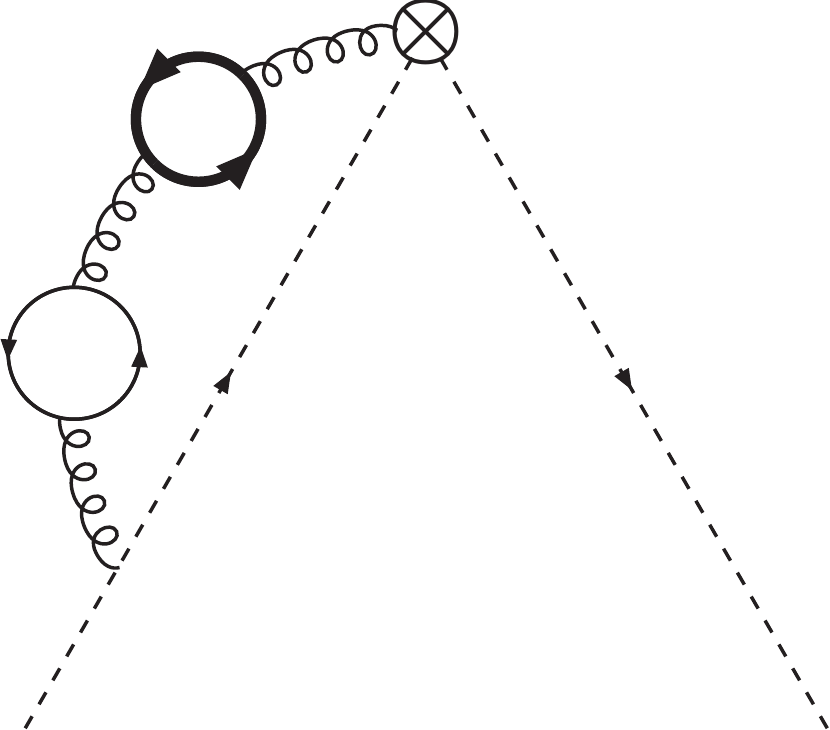}
\vspace*{-9mm}
\begin{center}
{\footnotesize (2)}
\end{center}
\end{minipage}
\hspace*{8mm}
\begin{minipage}[c]{0.20\linewidth}
  \includegraphics[width=1\textwidth]{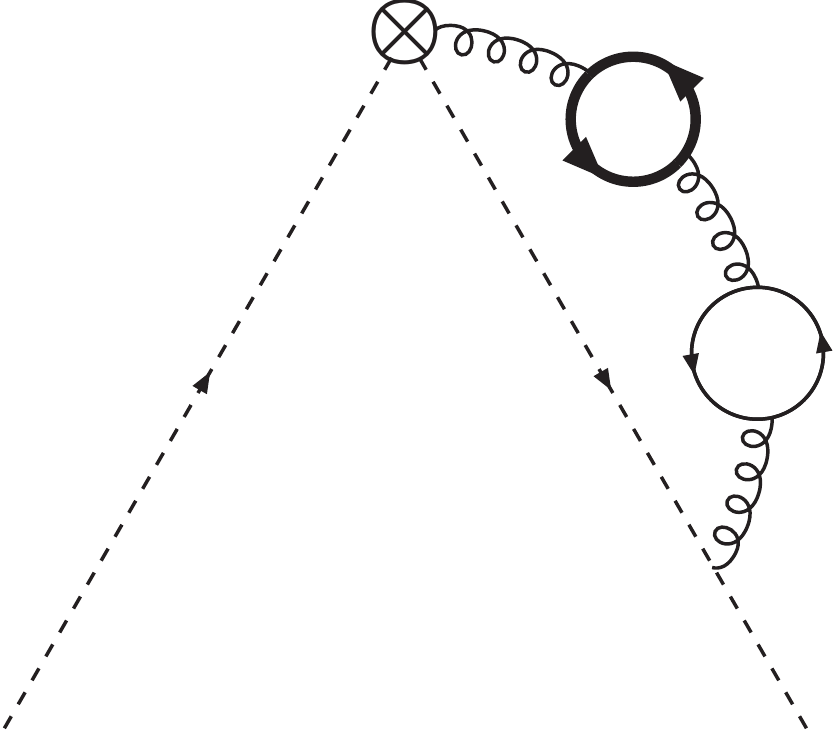}
\vspace*{-9mm}
\begin{center}
{\footnotesize (3)}
\end{center}
\end{minipage}
\caption{\small Diagrams for the two-mass contributions to $\tilde{A}_{qq,Q}^{(3), \rm NS}$. The curly lines denote
gluons, the dashed arrow 
line represents the external massless quarks, while the thick solid arrow line represents a quark of mass $m_1$, 
and the thin arrow line a quark of mass $m_2$. We assume $m_1 > m_2$. All diagrams have been drawn using 
{\tt Axodraw}  \cite{Vermaseren:1994je}.}
\label{NSdiagrams}
\end{center}
\end{figure}

In Figure \ref{NSdiagrams}, we show a sample of the diagrams contributing to $A_{qq,Q}^{(3), \rm NS}$. 
The remaining diagrams are related to these by the exchange $m_1 \leftrightarrow m_2$.
In the case of the diagrams 2 and 3 in Figure \ref{NSdiagrams},
the pre--factors $C_j\left(\ep,N\right)$ appearing in Eq.~(\ref{I4F3})
will contain a sum arising from the vertex operator Feynman rule
(see Section~8.1 of Ref.~\cite{Bierenbaum:2009mv}), which can be
evaluated in terms of single harmonic sums using the Mathematica packages {\tt Sigma}~\cite{SIG1,SIG2}, 
{\tt HarmonicSums}~\cite{HARMONICSUMS,Ablinger:2011te,Ablinger:2013cf},
{\tt EvaluateMultiSums} and {\tt SumProduction} \cite{EMSSP}.

For the constant part, $\tilde{a}_{qq,Q}^{(3),\rm{NS}}$, which is the only term in Eq. (\ref{AthhNS3}) that is
not determined by the renormalization prescription, we obtain the following result
\begin{eqnarray}
\tilde{a}_{qq,Q}^{(3), \rm{NS}} &=&
 C_F T_F^2 \Biggl\{
 \left(\frac{4}{9} S_1-\frac{3 N^2+3 N+2}{9 N (N+1)}\right) \Biggl[
-24 \big(L_1^3+L_2^3+\left(L_1 L_2+2 \zeta_2+5\right) \left(L_1+L_2\right)\big)
\nonumber\\&&
+\frac{\eta+1}{\eta^{3/2}} \left(5 \eta^2+22 \eta+5\right) \biggl(
-\frac{1}{4} \ln^2(\eta) \ln\left(\frac{1+\sqrt{\eta}}{1-\sqrt{\eta}}\right)
+2 \ln(\eta) \Li_2\left(\sqrt{\eta}\right)
-4 \Li_3\left(\sqrt{\eta}\right)
\biggr)
\nonumber\\&&
+\frac{\left(1+\sqrt{\eta}\right)^2}{2 \eta^{3/2}} \left(-10 \eta^{3/2}+5 \eta^2+42 \eta-10 \sqrt{\eta}+5\right) \big[
\Li_3\left(\eta\right)-\ln(\eta) \Li_2\left(\eta\right)\big]
+\frac{64}{3} \zeta_3
\nonumber\\&&
+\frac{8}{3} \ln^3(\eta)
-16 \ln^2(\eta) \ln(1-\eta)
+10 \frac{\eta^2-1}{\eta} \ln(\eta)
\Biggr]
+\frac{16 \left(405 \eta^2-3238 \eta+405\right)}{729 \eta} S_1
\nonumber\\&&
+\frac{4}{3} \left(\frac{3 N^4+6 N^3+47 N^2+20 N-12}{3 N^2 (N+1)^2}-\frac{40}{3} S_1+8 S_2\right) \left[\frac{4}{3} \zeta_2+(L_1+L_2)^2\right]
\nonumber\\&&
+\frac{8}{9} \left(\frac{130 N^4+84 N^3-62 N^2-16 N+24}{3 N^3 (N+1)^3}-\frac{52}{3} S_1+\frac{80}{3} S_2-16 S_3\right) \left(L_1+L_2\right)
\nonumber\\&&
+\biggl[-\frac{R_1}{18 N^2 (N+1)^2 \eta}
+\frac{2 \left(5 \eta^2+2 \eta+5\right)}{9 \eta} S_1
+\frac{32}{9} S_2\biggr] \ln^2(\eta)
-\frac{4 R_2}{729 N^4 (N+1)^4 \eta}
\nonumber\\&&
+\frac{3712}{81} S_2
-\frac{1280}{81} S_3
+\frac{256}{27} S_4
\Biggr\}~.
 \label{eq:aqqNSQ2M}
\end{eqnarray}
Here $S_{\vec{a}} \equiv S_{\vec{a}}(N)$ denote the (nested) harmonic sums \cite{HSUM}
\begin{eqnarray}
S_{b,\vec{a}}(N) = \sum_{k=1}^N \frac{({\rm sign}(b))^k}{k^{|b|}} S_{\vec{a}}(k),~~~S_\emptyset = 1,~~~b,a_i \in \mathbb{Z} 
\backslash \{0\}~.
\end{eqnarray}
The $R_i$'s are polynomials in $N$ and $\eta$, which we will not show here.

This result is symmetric under the exchange of masses $m_1 \leftrightarrow m_2$, and agrees with the results obtained
in \cite{Ablinger:2017err}\footnote{See also \cite{Ablinger:2011pb,Ablinger:2012qj}.}
for the individual fixed moments $N=2,4,6$. Eq. (\ref{eq:aqqNSQ2M}) vanishes for $N=1$ as
expected. A similar result is obtained in the case of the transversity contribution, where the change 
$\gamma_{qq}^{\rm NS} \rightarrow \gamma_{qq}^{\rm NS,trans}$ \cite{Blumlein:2009rg,Ablinger:2014vwa} and the corresponding 
one in the 2-loop OMEs must be performed in Eq.~(\ref{AthhNS3}), cf.~\cite{Ablinger:2017err}.

\begin{figure}[h]
\centering
\includegraphics[width=0.7\textwidth]{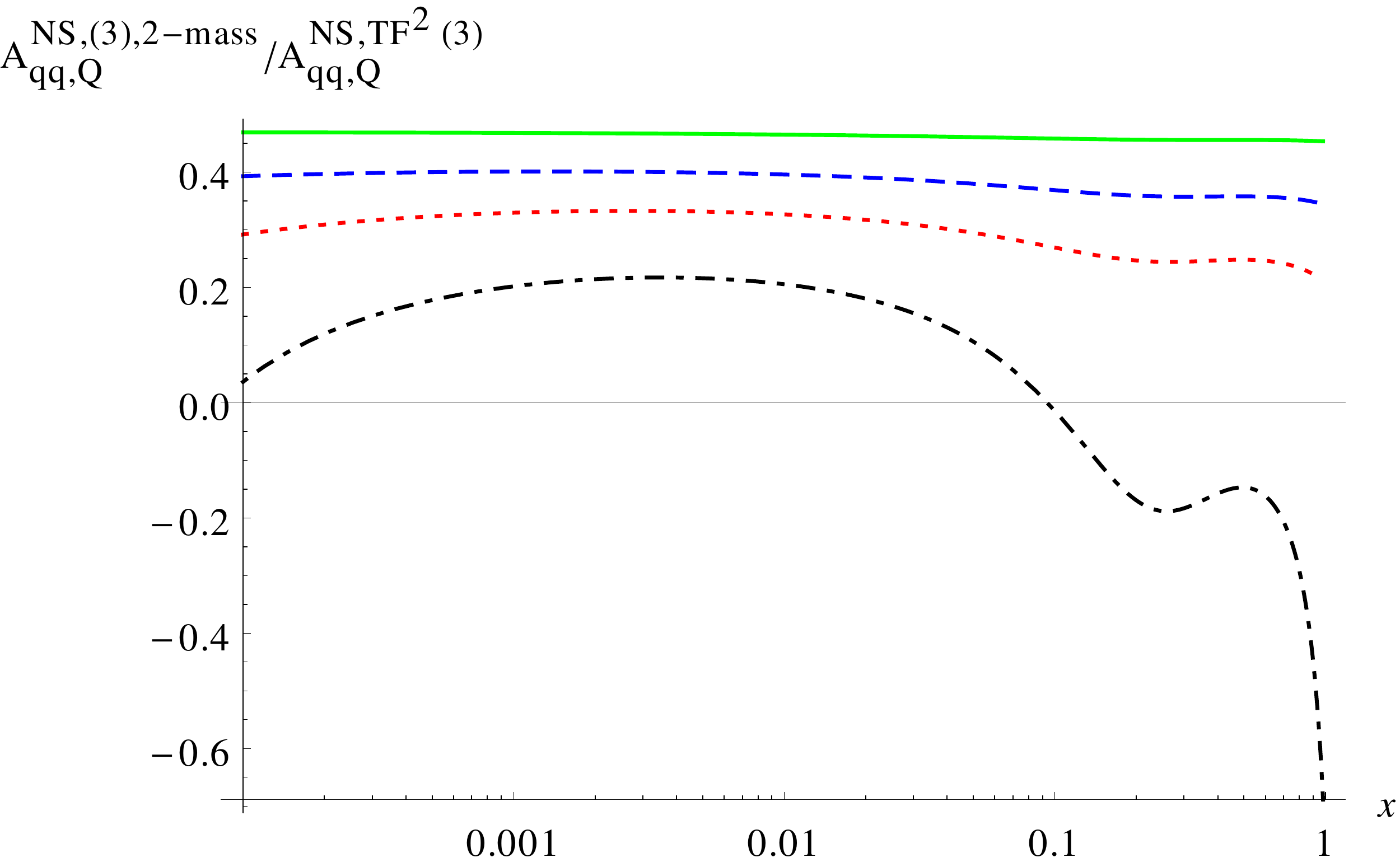}
\caption{\label{FIG:NS}
\small The ratio of the genuine 2-mass contributions to $A_{qq,Q}^{(3), \rm NS}$ to the complete $T_F^2$-part of
massive 3-loop OME $A_{qq,Q}^{(3), \rm NS}$ as a function of $x$ and $Q^2$, for $m_c = 1.59~\GeV, m_b = 4.78~\GeV$ 
in the on-shell scheme.
Dash-dotted line:  $\mu^2 =30~\GeV^2$;
Dotted line:  $\mu^2 =50~\GeV^2$;
Dashed line:  $\mu^2 =100~\GeV^2$;
Full line:  $\mu^2 =1000~\GeV^2$.
The single mass contributions are given in Ref.~\cite{Ablinger:2014vwa};~from Ref.~\cite{Ablinger:2017err}.}
\end{figure}

By a Mellin inversion of Eq. (\ref{eq:aqqNSQ2M}) we obtain the result in $x$-space \cite{Ablinger:2017err}.
In Figure \ref{FIG:NS}, we plot the ratio of the two--mass contribution to the complete $T_F^2$ term including 
the single mass results \cite{Ablinger:2014vwa} as a function of $x$ and $Q^2$, 
taking the masses of the heavy quarks to be those of the charm and bottom quarks in the on-shell scheme. 
The two--mass contributions become more important for larger values of $Q^2$, where the dependence on the ratio as a 
function of $x$
becomes almost flat approaching values of $O(0.4)$.

\subsection{The \boldmath $gq$-contribution}

\begin{figure}[h]
\begin{center}
\begin{minipage}[c]{0.20\linewidth}
  \includegraphics[width=1\textwidth]{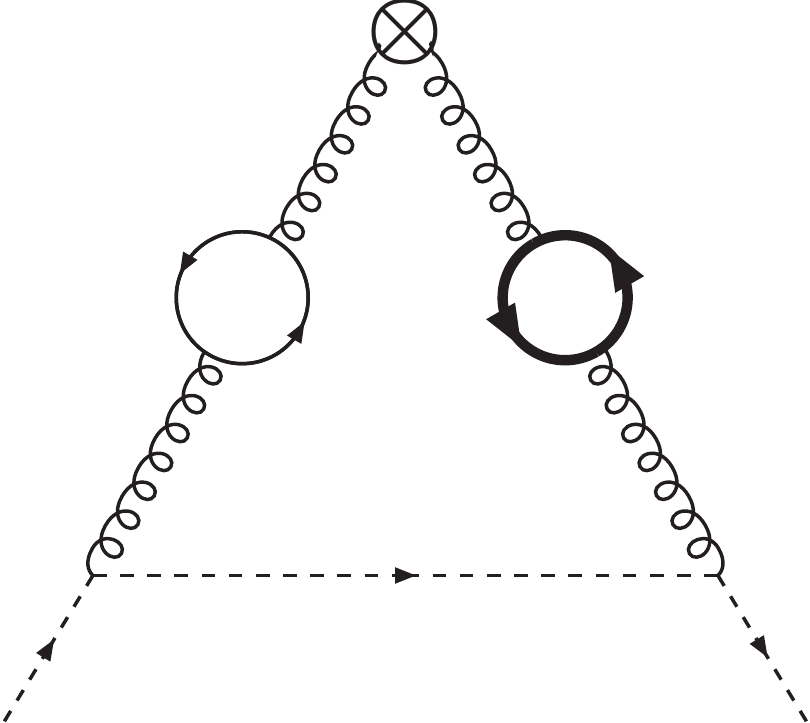}
\vspace*{-9mm}
\begin{center}
{\footnotesize (1)}
\end{center}
\end{minipage}
\hspace*{8mm}
\begin{minipage}[c]{0.20\linewidth}
  \includegraphics[width=1\textwidth]{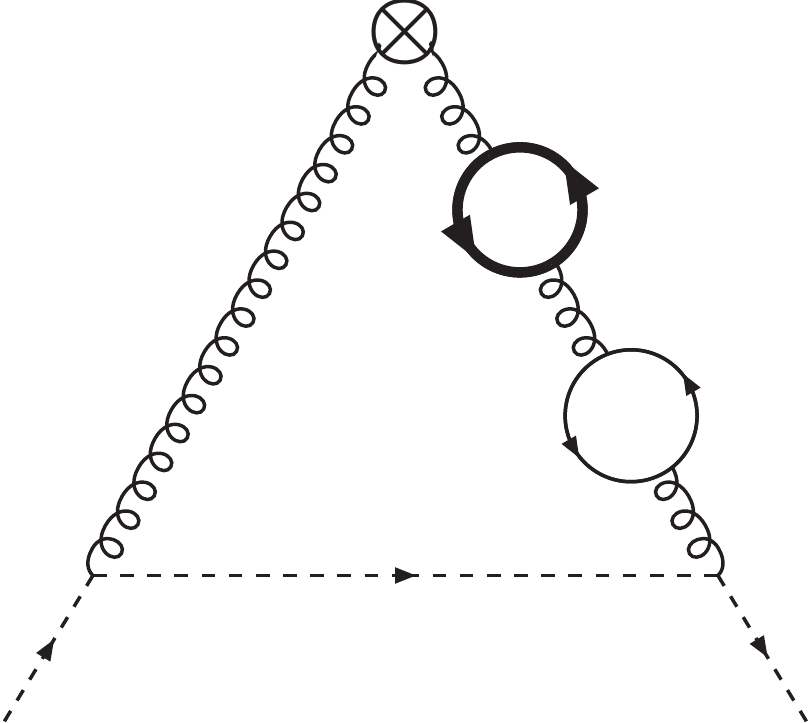}
\vspace*{-9mm}
\begin{center}
{\footnotesize (2)}
\end{center}
\end{minipage}
\hspace*{8mm}
\begin{minipage}[c]{0.20\linewidth}
  \includegraphics[width=1\textwidth]{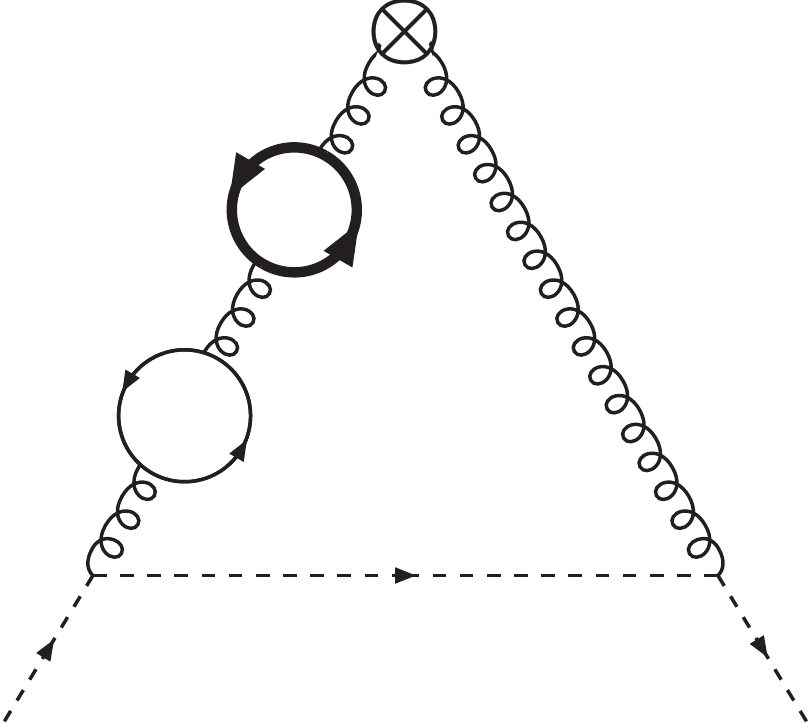}
\vspace*{-9mm}
\begin{center}
{\footnotesize (3)}
\end{center}
\end{minipage}
\caption{\small Diagrams for the two-mass contributions to $\tilde{A}_{gq,Q}^{(3)}$. The curly lines denote gluons, the 
dashed arrow line represents the external
massless quarks, while the thick solid arrow line represents a quark of mass $m_1$, and the thin arrow line a quark of mass $m_2$. We assume $m_1 > m_2$.}
\label{GQdiagrams}
\end{center}
\end{figure}

\vspace*{1mm}
\noindent
The genuine two-mass contributions to the OME $A_{gq,Q}^{(3)}$ can be calculated in a similar way as $A_{qq,Q}^{{\sf NS},(3)}$.
A sample of the contributing diagrams is shown in Figure \ref{GQdiagrams}. There are three more diagrams related to these by the exchange
$m_1 \leftrightarrow m_2$.

From the renormalization procedure, one obtains the following pole structure,
\begin{eqnarray}
\Athathat_{gq,Q}^{(3)}&=&
-\frac{16}{\ep^3} \gamma_{gq}^{(0)} \beta_{0,Q}^2
-\frac{4}{\ep^2} \Bigl[
3 \gamma_{gq}^{(0)} \beta_{0,Q}^2 \left(L_2+L_1\right)
+\beta_{0,Q} \hat{\gamma}_{gq}^{(1)}
\Bigr]
+\frac{1}{\ep} \biggl[
-6 \gamma_{gq}^{(0)} \beta_{0,Q}^2 
\left(L_2^2+L_1 L_2+L_1^2\right)
\NN\\&&
-3 \beta_{0,Q} \hat{\gamma}_{gq}^{(1)} \left(L_2+L_1\right)
+\frac{2}{3} \hat{\tilde{\gamma}}_{gq}^{(2)}
-12 a_{gq}^{(2)} \beta_{0,Q}
\biggr] 
+\tilde{a}_{gq,Q}^{(3)}
  ~, \label{AhhhgqQ3}
\end{eqnarray}
and the renormalized operator matrix element in the $\MS$--scheme reads
\begin{eqnarray}
 \tilde{A}_{gq,Q}^{(3), \MS}&=&
\gamma_{gq}^{(0)} \beta_{0,Q}^2 \left(2 L_2^3+2 L_1^3 + \frac{3}{2} L_2^2 L_1+\frac{3}{2} L_1^2 L_2\right) 
+\frac{3}{2} \beta_{0,Q} \hat{\gamma}_{gq}^{(1)} \left(L_2^2+L_1^2\right)
\NN\\&&
+\biggl\{
6 a_{gq}^{(2)} \beta_{0,Q}
+\frac{3}{2} \gamma_{gq}^{(0)} \beta_{0,Q}^2 \zeta_2
\biggr\} \left(L_2+L_1\right)
+12 \overline{a}_{gq}^{(2)} \beta_{0,Q}
+\tilde{a}_{gq,Q}^{(3)}
~. \label{Agq3QMSren}
\nonumber\\
\end{eqnarray}

One obtains the following result for the constant part of the two--mass unrenormalized OME,
\begin{eqnarray}
\tilde{a}_{gq,Q}^{(3)}&=&C_F T_F^2 \frac{1+(-1)^N}{2} \Biggl\{
p_{gq}^{(0)} \biggl[
16 \biggl(L_1^3+L_2^3+\left(L_1 L_2+2 \zeta_2+\frac{26}{3}\right) (L_1+L_2)\biggr)
\nonumber\\&&
-\frac{4}{3 \eta^{3/2}} \left(
\big(1+\sqrt{\eta }\big)^2 R_8  \Li_3(-\sqrt{\eta})
-\big(1-\sqrt{\eta }\big)^2 R_9 \Li_3(\sqrt{\eta})
\right)
-\frac{16}{9} \ln^3(\eta)
\nonumber\\&&
+\biggl(
\frac{2 \big(1+\sqrt{\eta }\big)^2}{3 \eta^{3/2}} R_8 \Li_2(-\sqrt{\eta})
-\frac{2 \big(1-\sqrt{\eta }\big)^2}{3 \eta^{3/2}} R_9  \Li_2(\sqrt{\eta})
-\frac{20}{3 \eta} \left(\eta^2-1\right)
\biggr) \ln(\eta)
\nonumber\\&&
+\biggl(
\frac{\big(1+\sqrt{\eta }\big)^2}{6 \eta^{3/2}} R_8 \ln(1+\sqrt{\eta})
-\frac{\big(1-\sqrt{\eta }\big)^2}{6 \eta^{3/2}} R_9  \ln(1-\sqrt{\eta})
-\frac{16}{3} S_1
\biggr) \ln^2(\eta)
\nonumber\\&&
-\frac{64}{27} S_1^3
-\frac{128}{27} S_3
-\frac{64}{3} \left(\zeta_2+\frac{1}{3} S_2\right) S_1
-\frac{128}{9} \zeta_3
\biggr]
-\frac{R_{10} \ln^2(\eta)}{3 \eta (N-1) N (N+1)^2}
\nonumber\\&&
+16 \left[-\frac{1}{(N+1)^2}+p_{gq}^{(0)} \left(\frac{8}{3}-S_1\right)\right]\left((L_1+L_2)^2-\frac{4}{3} (L_1+L_2) 
S_1\right)
\nonumber\\&&
+\left[\frac{32}{3} p_{gq}^{(0)} \left(S_2-S_1^2\right)-\frac{64 (8 N+5)}{9 (N+1)^3} \right] (L_1+L_2)
-\frac{64 R_{11} S_1}{27 (N-1) N (N+1)^3}
\nonumber\\&&
+\frac{64 \big(8 N^3+13 N^2+27 N+16\big)}{27 (N-1) N (N+1)^2} \left(
S_1^2
+S_2
+3 \zeta_2
\right)
-\frac{8 R_{12}}{243 \eta (N-1) N (N+1)^4}
\Biggr\},
\nonumber\\&&
\label{eq:agq}
\end{eqnarray}
where $p_{gq}^{(0)}$ is the color--stripped 1--loop anomalous dimension,
\begin{eqnarray}
p_{gq}^{(0)} = \frac{2+N+N^2}{(N-1)N(N+1)}.
\end{eqnarray}
Again, we will omit the explicit expressions of the polynomials $R_i$.

The corresponding $x$-space expression is given in the appendix of Ref.~\cite{Ablinger:2017err}. 
Again, Eq. (\ref{eq:agq}) agrees with previously calculated fixed moments for $N=2,4,6$. 
The behaviour of the ratio of the genuine 2-mass contribution to the complete $T_F^2$
3-loop result for $A_{gq,Q}^{(3)}$ at large values of $Q^2$ is similar to the one exhibited
in the non-singlet case in Figure \ref{FIG:NS}.

\section{The gluonic operator matrix element $\tilde{A}_{gg}^{(3)}$}
\label{GGsection}

\vspace*{1mm}
\noindent
The three--loop contributions to $\tilde{A}_{gg}$ have no effect on the DIS structure functions at NNLO, since only the
corresponding two--loop contributions appear in the expressions for the massive Wilson coefficients (see Eq.s 
(2.20)--(2.24) of Ref.~\cite{Ablinger:2017err}). They are, however, needed in
order to obtain the gluonic transition relation of the VFNS at three loops. The calculation of $\tilde{A}_{gg}^{(3)}$ is considerably more elaborate
than those of the previous OMEs, and leads to new types of functions, namely, generalized sums in $N$-space and 
the generalized harmonic polylogarithms with square root letters in the alphabet in $x$-space.

\begin{figure}[ht]
\begin{center}
\begin{minipage}[c]{0.20\linewidth}
  \includegraphics[width=1\textwidth]{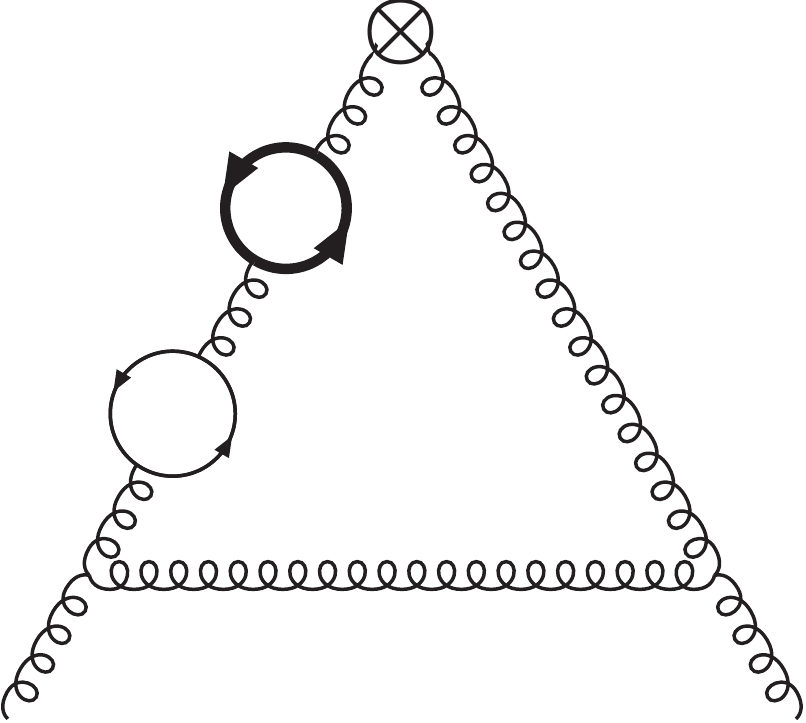}
\vspace*{-11mm}
\begin{center}
{\footnotesize (1)}
\end{center}
\end{minipage}
\hspace*{1mm}
\begin{minipage}[c]{0.20\linewidth}
  \includegraphics[width=1\textwidth]{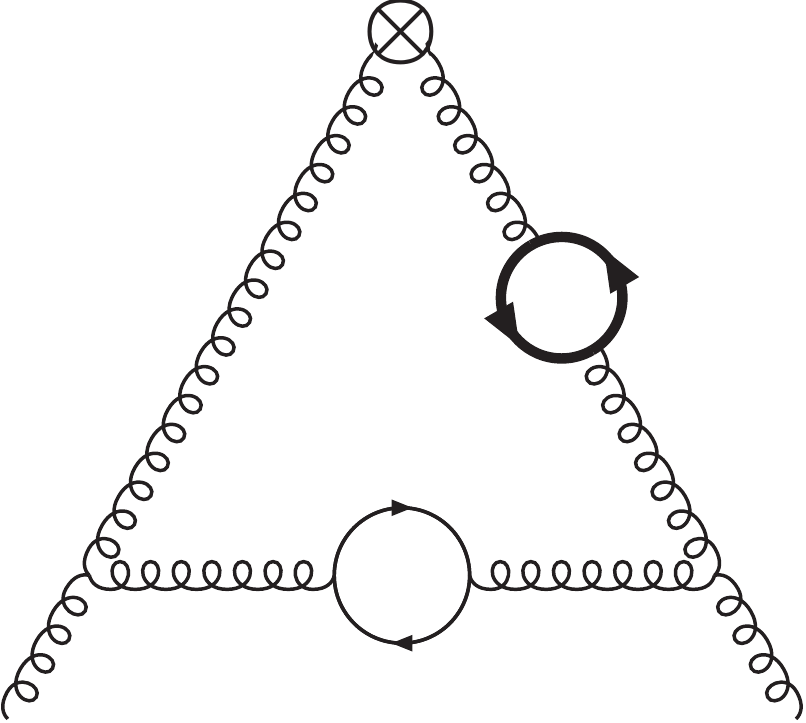}
\vspace*{-11mm}
\begin{center}
{\footnotesize (2)}
\end{center}
\end{minipage}
\hspace*{1mm}
\begin{minipage}[c]{0.20\linewidth}
  \includegraphics[width=1\textwidth]{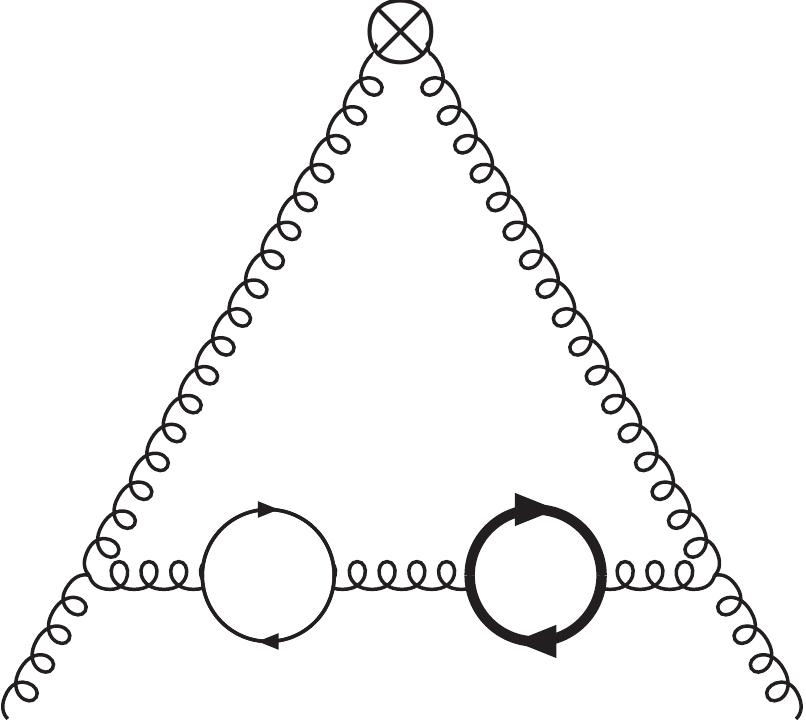}
\vspace*{-11mm}
\begin{center}
{\footnotesize (3)}
\end{center}
\end{minipage}
\hspace*{1mm}
\begin{minipage}[c]{0.20\linewidth}
  \includegraphics[width=1\textwidth]{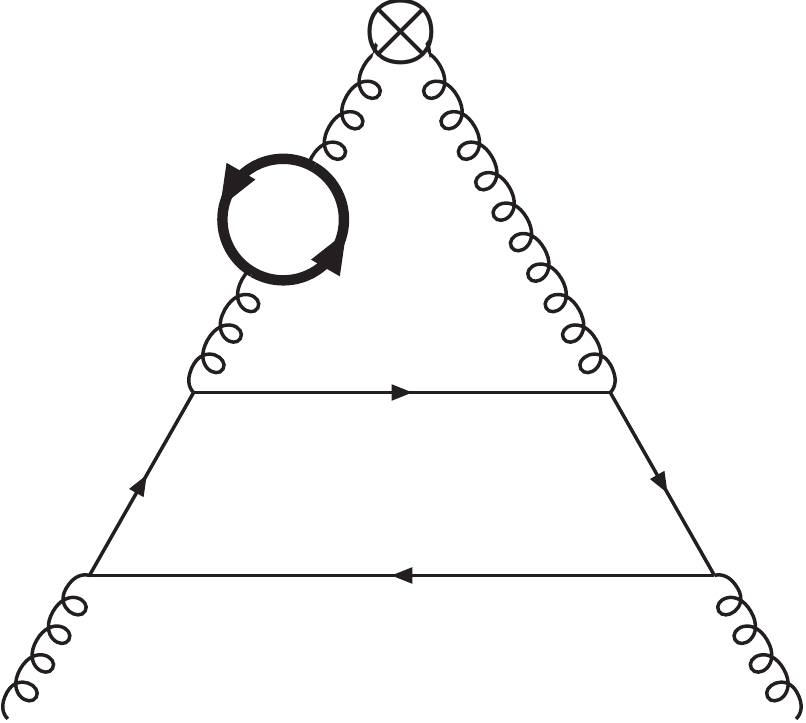}
\vspace*{-11mm}
\begin{center}
{\footnotesize (4)}
\end{center}
\end{minipage}

\vspace*{5mm}

\begin{minipage}[c]{0.20\linewidth}
  \includegraphics[width=1\textwidth]{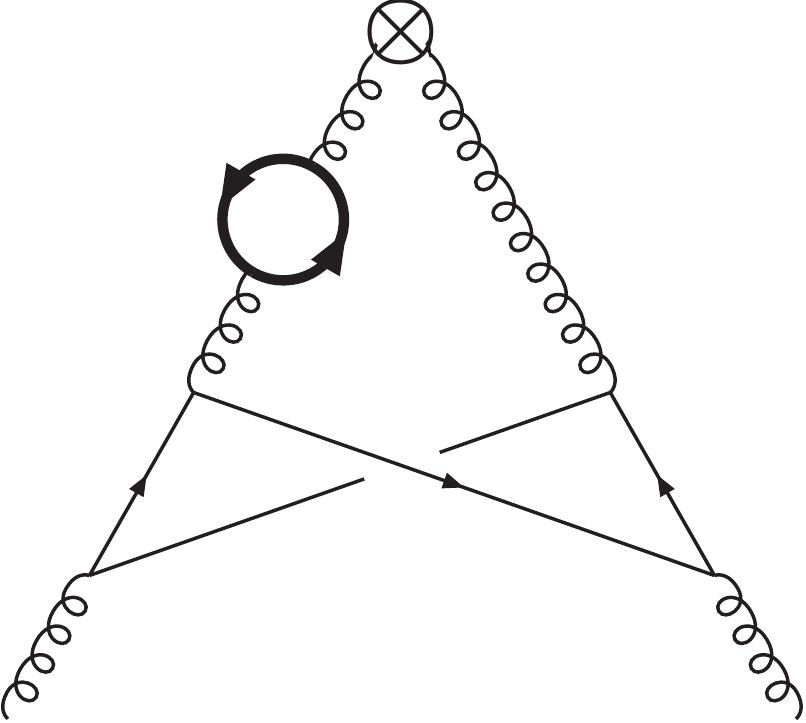}
\vspace*{-11mm}
\begin{center}
{\footnotesize (5)}
\end{center}
\end{minipage}
\hspace*{1mm}
\begin{minipage}[c]{0.20\linewidth}
  \includegraphics[width=1\textwidth]{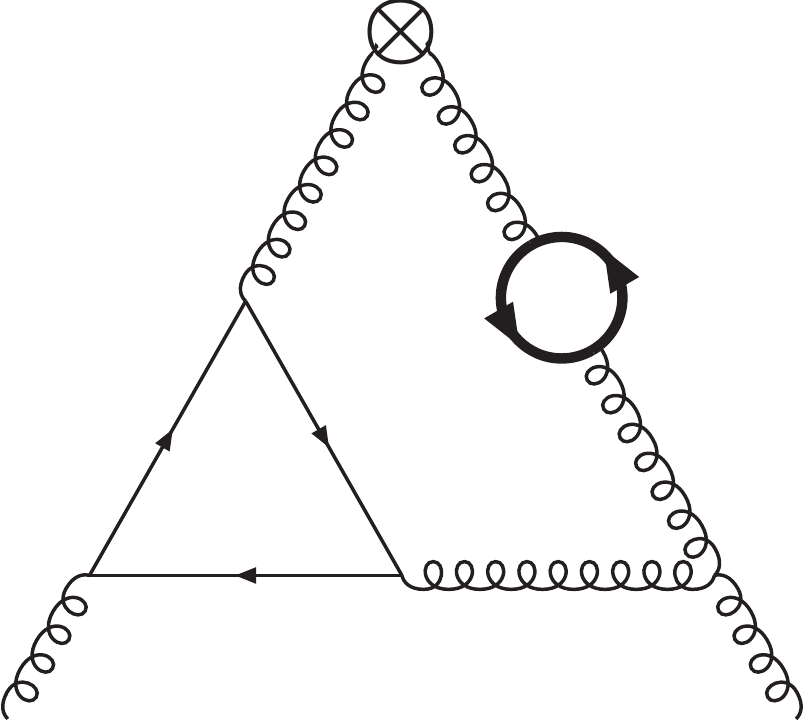}
\vspace*{-11mm}
\begin{center}
{\footnotesize (6)}
\end{center}
\end{minipage}
\hspace*{1mm}
\begin{minipage}[c]{0.20\linewidth}
  \includegraphics[width=1\textwidth]{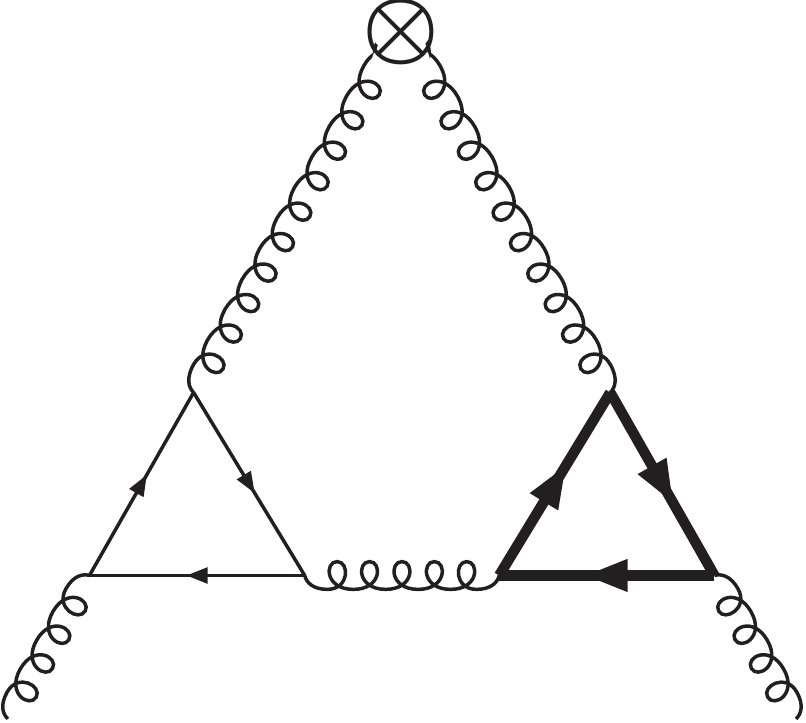}
\vspace*{-11mm}
\begin{center}
{\footnotesize (7)}
\end{center}
\end{minipage}
\hspace*{1mm}
\begin{minipage}[c]{0.20\linewidth}
  \includegraphics[width=1\textwidth]{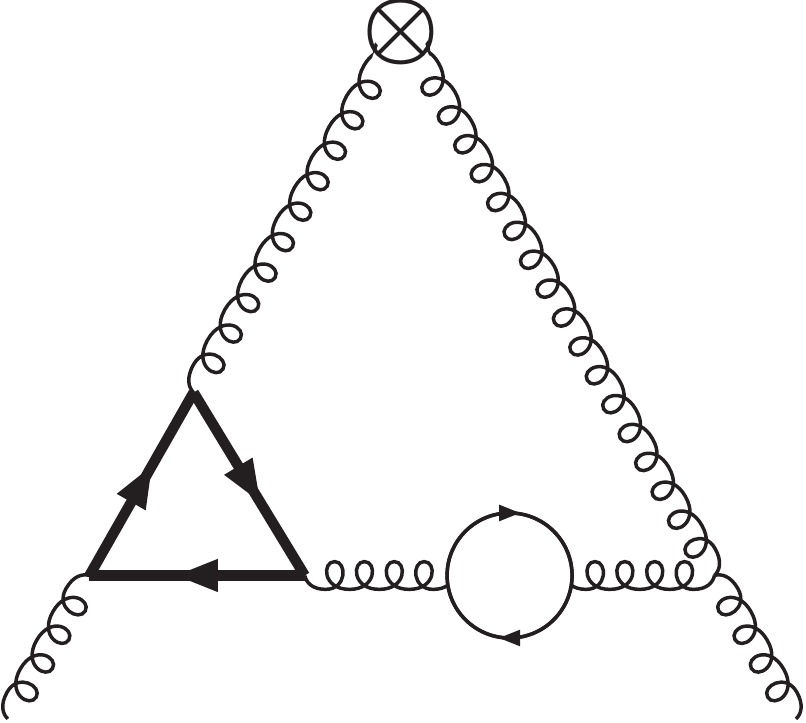}
\vspace*{-11mm}
\begin{center}
{\footnotesize (8)}
\end{center}
\end{minipage}

\vspace*{5mm}

\begin{minipage}[c]{0.20\linewidth}
  \includegraphics[width=1\textwidth]{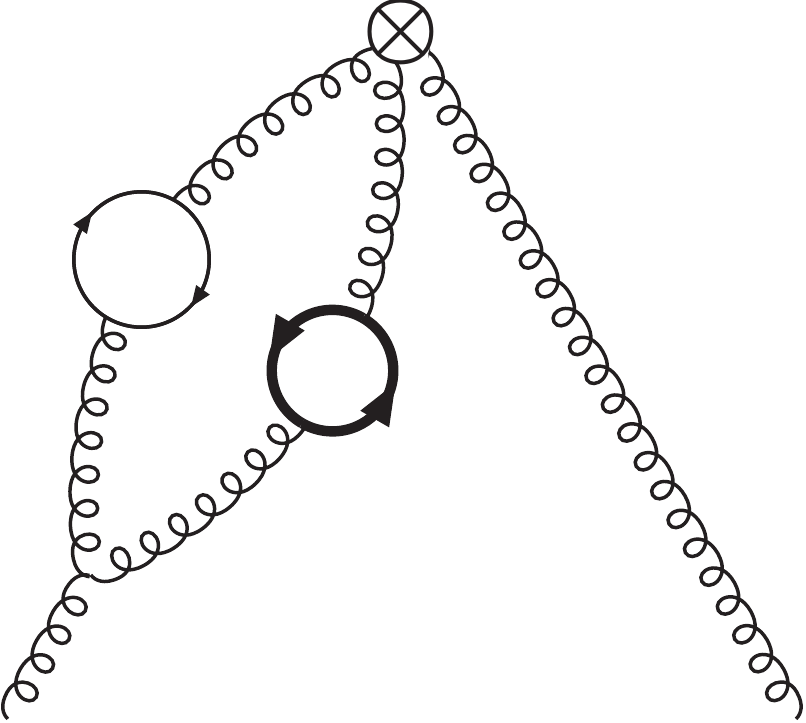}
\vspace*{-11mm}
\begin{center}
{\footnotesize (9)}
\end{center}
\end{minipage}
\hspace*{1mm}
\begin{minipage}[c]{0.20\linewidth}
  \includegraphics[width=1\textwidth]{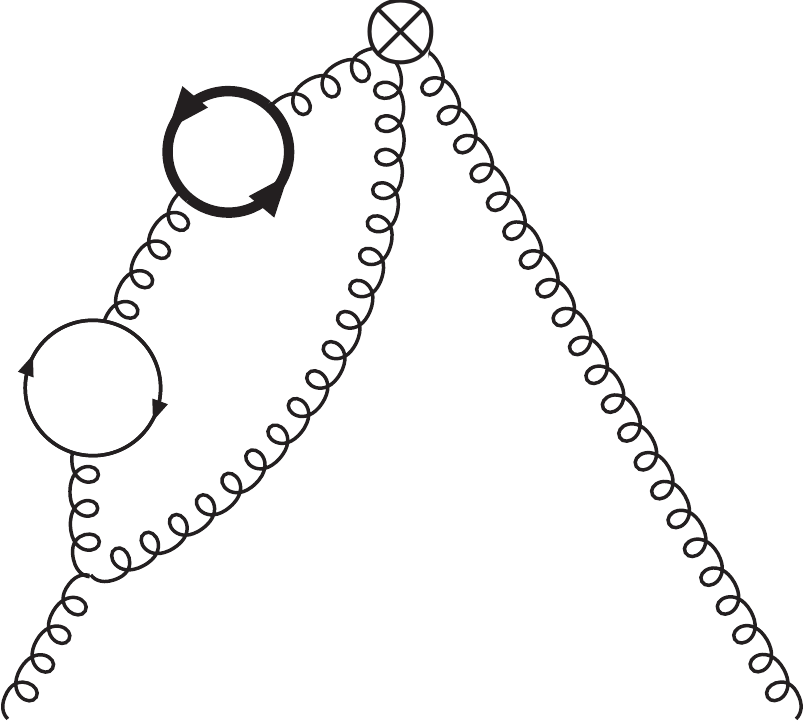}
\vspace*{-11mm}
\begin{center}
{\footnotesize (10)}
\end{center}
\end{minipage}
\hspace*{1mm}
\begin{minipage}[c]{0.20\linewidth}
  \includegraphics[width=1\textwidth]{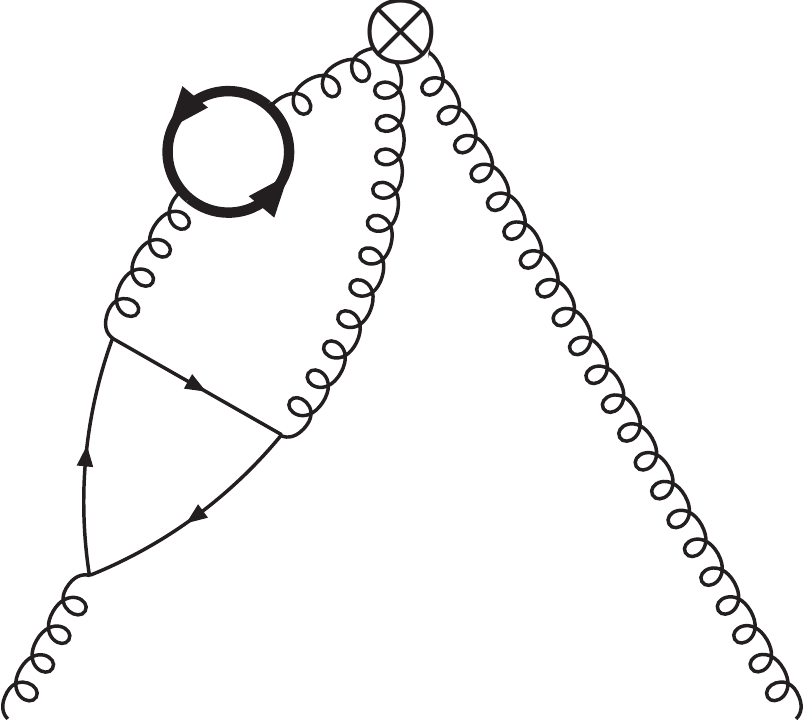}
\vspace*{-11mm}
\begin{center}
{\footnotesize (11)}
\end{center}
\end{minipage}
\hspace*{1mm}
\begin{minipage}[c]{0.20\linewidth}
  \includegraphics[width=1\textwidth]{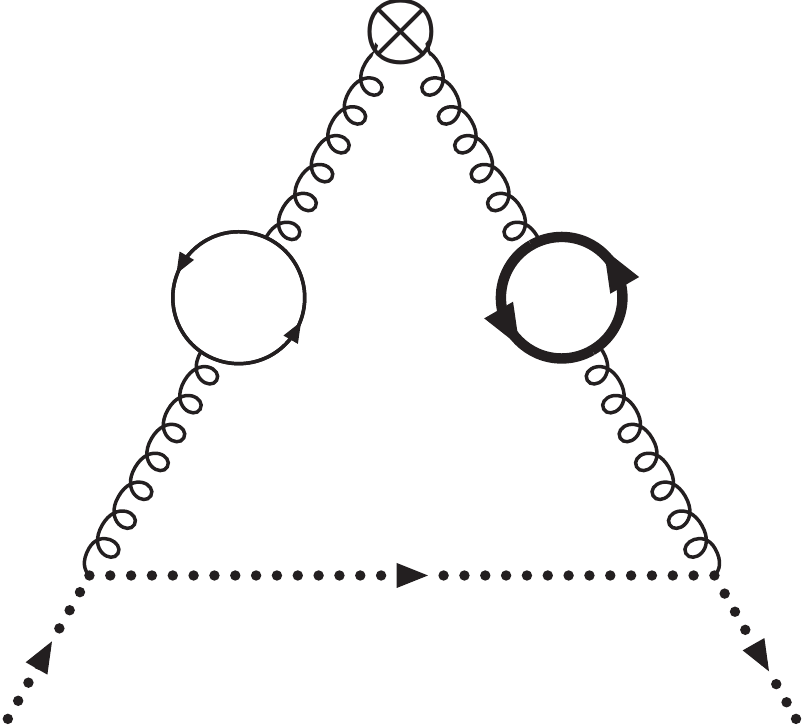}
\vspace*{-11mm}
\begin{center}
{\footnotesize (12)}
\end{center}
\end{minipage}
\caption{\small Independent diagrams for the two-mass contributions to $\tilde{A}_{gg}^{(3)}$. The curly lines
denote gluons, the dotted line represents a ghost, the thick solid arrow line represents a quark of mass $m_1$, 
and the thin arrow line a quark of mass $m_2$. We assume $m_1 > m_2$.}
\label{GGdiagrams}
\end{center}
\end{figure}

There is a total of 76 irreducible diagrams contributing to the two--mass part of $A_{gg}^{(3)}$, including six diagrams
with a ghost in the external lines. Only twelve of them, shown in Figure \ref{GGdiagrams}, are truly independent. All other
diagrams are related to these either by symmetry, the exchange of masses or reversal of a fermion line.

In Ref.~\cite{Ablinger:2017err}, we presented the results for the scalar versions of the first eight diagrams of Figure 
\ref{GGdiagrams}. 
We have now completed the calculations of all but two of the diagrams in the physical case. The details of how this calculation has been performed
will be explained in a future publication. As a preview, we show here the result for one of the diagrams, namely diagram 6 
of Figure \ref{GGdiagrams}.
We get,

\begin{eqnarray}
D_6^B \left( N \right) &=& C_A T_F^2 \frac{ 1 + (-1)^N }{2} \Biggl\{ 
\frac{16 \big( 29 N^3+41 N^2+47 N+47 \big) }{27 (N-1) N (N+1)^2 \ep^3}  
+ \frac{1}{\ep^2} \Biggl[
\frac{8 P_ 5}{81 (N-1)^2 N^2 (N+1)^3} 
\NN \\&&
-\frac{8 \big( 13 N^3+21 N^2+23 N+27 \big) }{ 9 (N-1) N (N+1)^2} H_ 0(\eta)
-\frac{8 \big( 23 N^3+19 N^2+21 N+13\big) }{27(N-1) N (N+1)^2} S_ 1
\Biggr] 
\NN \\&&
+\frac{1}{\ep}
\Biggl[
\frac{4 P_ 6}{81 (N-1)^3 N^3 (N+1)^4}
+\frac{2 \big( 17 N^3-3 N^2-5 N-21\big) }{27(N-1) N (N+1)^2} S_ 2
\NN \\&&
+\frac{2\big(23 N^3+43 N^2+45 N+61\big) }{9(N-1) N (N+1)^2} H^2_0(\eta) 
+\frac{2 \big(29 N^3+41 N^2+47 N+47\big) }{27(N-1) N (N+1)^2} S_ 1^2
\NN \\&&
-\frac{4 P_3}{27(N-1)^2 N^2 (N+1)^3} H_ 0(\eta) 
+\frac{2  \big(29 N^3+41 N^2+47 N+47\big) }{9(N-1) N (N+1)^2} \zeta_2
\NN \\&&
+\biggl(
  -\frac{4 P_ 4}{81 (N-1)^2 N^2 (N+1)^3}
  +\frac{4 \big(7 N^3-N^2-3 N-7\big) }{9(N-1) N (N+1)^2} H_ 0(\eta) 
\biggr) S_ 1
\Biggr] 
\NN \\&&
-\frac{ 53 N^3+109 N^2+111 N+163 }{27 (N-1) N (N+1)^2} H_0^3(\eta)
-\frac{ 23 N^3+19 N^2+21 N+13 }{81(N-1) N (N+1)^2} S_1^3 
\NN \\&&
+\frac{5 N^3+11 N^2+11 N+17}{9(N-1) N (N+1)^2} \Bigl(4 H_ 0(\eta) H_{0,1}(\eta)-4 H_{0,0,1}(\eta)-2 H_0^2(\eta) H_ 1(\eta)\Bigr)
%
\NN \\&&
+\frac{4 P_8}{45 \eta (N-1) N (N+1)^2} 
\biggl\{
 H_ 0(\eta) S_ {1,1}\left({{\frac{\eta }{\eta -1},\frac{\eta -1}{\eta }},N}\right) 
- S_ {1,2}\left({{\frac{\eta - 1}{\eta },\frac{\eta }{\eta -1}},N}\right)
\NN \\&&
+ S_ {1,2}\left({{\frac{\eta }{\eta - 1 },\frac{\eta - 1 }{\eta }},N}\right)
+ S_ {1,1,1}\left({{\frac{\eta - 1 }{\eta },1,\frac{\eta }{\eta - 1 }},N}\right)
+ S_ {1,1,1}\left({{\frac{\eta - 1 }{\eta },\frac{\eta }{\eta - 1},1},N}\right)
\biggr\}
\NN \\&&
-\frac{2 P_9}{405 (N-1) N (N+1)^2 \eta } S_3 
+\frac{P_{14}}{3645 (N-1)^4 N^4 (N+1)^5 (2 N-5) (2 N-3) (2 N-1) \eta } 
\NN \\&&
+\frac{P_{15}}{90 (N-1) N (N+1) \eta^{3/2}} 
\Bigl(
 H_ 0(\eta) H_ {0,-1}\big(\sqrt{\eta }\big)
+H_ 0(\eta) H_ {0,1}\big(\sqrt{\eta }\big) 
-\frac{1}{4} H_ 0(\eta)^2 H_ 1\big(\sqrt{\eta }\big) 
\NN \\&&
-2 H_ {0,0,-1}\big(\sqrt{\eta }\big) 
-2 H_ {0,0,1}\big(\sqrt{\eta }\big)
\Bigr)
+\frac{P_ 5}{27 (N-1)^2 N^2 (N+1)^3} \zeta_2
\NN \\&&
+\frac{2^{-2 N-3} \binom{2 N}{N} P_{16} }{135 (N-1) N (N+1)^2 (2 N-5) (2 N-3) (2 N-1) \eta^2} 
\Biggl[
 \sum_{i=1}^N \frac{2^{2 i}}{\binom{2 i}{i} i^3}
 - \sum_{i=1}^N \frac{2^{2 i} S_ 1\left({i}\right)}{\binom{2 i}{i} i^2}
\NN \\&&
 + H_ 0(\eta) \sum_{i=1}^N \frac{2^{2 i}}{\binom{2 i}{i} i^2}
 + \sum_{i=1}^N \frac{2^{2 i} \big(\eta - 1 \big)^{-i} \eta^{i}}{\binom{2 i}{i} i} S_ {1,1}\left({{\frac{\eta - 1}{\eta },1},i}\right)
\NN \\&&
 + \frac{1}{2} H_0^2(\eta) \sum_{i=1}^N \frac{2^{2 i} \big(\eta - 1 \big)^{-i} \eta^{i}}{\binom{2 i}{i} i}
 - H_ 0(\eta) \sum_{i=1}^N \frac{2^{2 i} \big(\eta - 1 \big)^{-i} \eta^{i} }{\binom{2 i}{i} i} S_ 1\left({{\frac{\eta -1}{\eta }},i}\right)
\NN \\&&
 - \sum_{i=1}^N \frac{2^{2 i} \big(\eta -1 \big)^{-i} \eta^{i} }{\binom{2 i}{i} i} S_ 2\left({{\frac{\eta -1}{\eta }},i}\right)
\Biggr]
-\frac{2 \big(11 N^3+47 N^2+41 N+89\big) }{27(N-1) N (N+1)^2} \zeta_3
\NN \\&&
+ \left( \frac{\eta}{\eta-1} \right)^N \frac{P_{11}}{540(N-1)^2 N^2 (N+1)^2 (2 N-5) (2 N-3) (2 N-1) \eta}
\biggl[
S_ 2\left({{\frac{\eta - 1}{\eta }},N}\right)
\NN \\&&
  +H_ 0(\eta) S_ 1\left({{\frac{\eta - 1}{\eta }},N}\right) 
  -\frac{1}{2} H_0^2(\eta)
  - S_ {1,1}\left({{\frac{\eta - 1}{\eta },1},N}\right) 
\biggr]
\NN \\&&
+ \frac{2^{-2 N-4} \binom{2 N}{N} P_{16} }{135(N-1) N (N+1)^2 (2 N-5) (2 N-3) (2 N-1) \eta^{3/2}}
\Bigl[
8 H_ {0,0,-1}\big(\sqrt{\eta }\big) 
\NN \\&&
  +\bigl(
H_ {-1}\big(\sqrt{\eta }\big)
  +H_ 1\big(\sqrt{\eta }\big) 
  \bigr)  H_0^2(\eta)
  + 8 H_ {0,0,1}\big(\sqrt{\eta }\big)
  - 4 H_ 0(\eta) \bigl(
H_ {0,-1}\big(\sqrt{\eta }\big) 
  +H_ {0,1}\big(\sqrt{\eta }\big)
  \bigr)
\Bigr]
\NN \\&&
+\biggl[
  -\frac{35 N^3+63 N^2+73 N+81}{27(N-1) N (N+1)^2} S_ 2 
  +\frac{-5 N^3+23 N^2+33 N+41 }{9(N-1) N (N+1)^2} H_0^2(\eta) 
\NN \\&&
  +\frac{2 P_1}{27 (N-1)^2 N^2 (N+1)^3} H_ 0(\eta)
  -\frac{23 N^3+19 N^2+21 N+13}{9(N-1) N (N+1)^2} \zeta_2
\NN \\&&
  +\frac{P_{12}}{810(N-1)^3 N^3 (N+1)^4 (2 N-5) (2 N-3) (2 N-1) \eta }
\biggr] S_ 1
\NN \\&&
+\biggl[
  \frac{P_5}{81 (N-1)^2 N^2 (N+1)^3}
  -\frac{13 N^3+21 N^2+23 N+27}{9(N-1) N (N+1)^2} H_ 0(\eta) 
\biggr] S_1^2
\NN \\&&
+\biggl[
  \frac{P_ 2}{81 (N-1)^2 N^2 (N+1)^3}
  +\frac{P_7}{45 (N-1) N (N+1)^2 \eta } H_ 0(\eta) 
\biggr] S_2
\NN \\&&
- \frac{2 P_8}{45 (N-1) N (N+1)^2 \eta } \biggl[
H_0^2(\eta) 
  + 2 S_ {1,1}\left({{\frac{\eta - 1 }{\eta },1},N}\right)
\biggr] S_ 1\left({{\frac{\eta }{\eta - 1}},N}\right)
\NN \\&&
+\biggl[
\frac{P_{13}}{810(N-1)^3 N^3 (N+1)^4 (2 N-5) (2 N-3) (2 N-1) \eta}
\NN \\&&
  -\frac{13 N^3+21 N^2+23 N+27}{3(N-1) N (N+1)^2}  \zeta_2
\biggr] H_ 0(\eta)
+\biggl[
  \frac{P_{10}}{540 (N-1)^2 N^2 (N+1)^3 \eta } 
\NN \\&&
  -\frac{P_{15}}{360 (N-1) N (N+1) \eta^{3/2}} H_ {-1}\big(\sqrt{\eta }\big)
\biggr] H_0^2(\eta)
\Biggr\},
\end{eqnarray}
where the $P_i$'s are polynomials in $N$ and $\eta$, and 
\begin{eqnarray}
S_{b,\vec{a}}(y,\vec{x},N) = \sum_{k=1}^N \frac{y^k}{k^b} S_{\vec{a}}(\vec{x},k),~~~S_\emptyset = 1,~~~b,a_i \in \mathbb{Z} 
\backslash \{0\},~~y, x_i \in \mathbb{C}
\end{eqnarray}
are generalized harmonic sums \cite{Ablinger:2013cf}.
The Mellin inversion of this result can be performed using the 
package {\tt HarmonicSums}.   
The results are given in terms of generalized iterated integrals defined by
\begin{equation}
G\left(\left\{f_1(\tau),f_2(\tau),\cdots,f_n(\tau)\right\},z\right)
=\int_0^z  d\tau_1~f_1(\tau_1)  
G\left(\left\{f_2(\tau),\cdots,f_n(\tau)\right\},\tau_1\right),
\label{Gfunctions}
\end{equation}
with
\begin{equation}
 G\Biggl(\Biggl\{\underbrace{\frac{1}{\tau},\frac{1}{\tau},
  \cdots,\frac{1}{\tau}}_{\text{n times}}\Biggr\},z\Biggr)
\equiv
\frac{1}{n!} \ln^n(z)~.
\end{equation}
These will also appear in the case of the pure singlet OME discussed in the next section. The letters in the
alphabet of these iterated integrals, i.e., the functions $f_i(\tau)$ in Eq. (\ref{Gfunctions}), may in general
depend on $\eta$.
\section{The pure singlet OME}
\label{PSsection}

\vspace*{1mm}
\noindent
In the case of the pure singlet two--mass contribution, the generic pole structure is given by
\begin{eqnarray}
\Athathat_{Qq}^{(3),\rm{PS}} &=&
\frac{16}{3 \ep^3} \gamma_{gq}^{(0)} \hat{\gamma}_{qg}^{(0)} \beta_{0,Q}
+\frac{1}{\ep^2} \biggl[
4 \gamma_{gq}^{(0)} \hat{\gamma}_{qg}^{(0)} \beta_{0,Q} \left(L_1+L_2\right)
+\frac{2}{3} \hat{\gamma}_{qg}^{(0)} \hat{\gamma}_{gq}^{(1)}
-\frac{8}{3} \beta_{0,Q} \hat{\gamma}_{qq}^{\rm{PS},(1)}
\biggr] 
\NN\\&&
+\frac{1}{\ep}
\biggl[
2 \gamma_{gq}^{(0)} \hat{\gamma}_{qg}^{(0)} \beta_{0,Q} \left(L_1^2+L_1 L_2+L_2^2\right)
+\biggl\{\frac{1}{2} \hat{\gamma}_{qg}^{(0)} \hat{\gamma}_{gq}^{(1)}
-2 \beta_{0,Q} \hat{\gamma}_{qq}^{\rm{PS},(1)}
\biggr\} \left(L_2+L_1\right)
\NN\\&&
+\frac{2}{3} \hat{\tilde{\gamma}}_{qq}^{(2),\rm{PS}}
-8 a_{Qq}^{(2),\rm{PS}} \beta_{0,Q}
+2 \hat{\gamma}_{qg}^{(0)} a_{gq}^{(2)}
\biggr] 
+\tilde{a}_{Qq}^{(3),\rm{PS}}
 \label{Ahhhqq3PSQ}~.
\end{eqnarray}
In the ${\MS}$--scheme, one obtains the following renormalized expression
\begin{eqnarray}
\tilde{A}_{Qq}^{(3), \MS, \rm{PS}} &=&
-\frac{1}{2} \gamma_{gq}^{(0)} \hat{\gamma}_{qg}^{(0)} \beta_{0,Q} 
\left(L_2^2 L_1+L_1^2 L_2+ \frac{4}{3} L_1^3+\frac{4}{3} L_2^3\right)
-\biggl\{
\frac{1}{4} \hat{\gamma}_{qg}^{(0)} \hat{\gamma}_{gq}^{(1)}
-\beta_{0,Q} \hat{\gamma}_{qq}^{\rm{PS},(1)}
\biggr\}
\left(L_2^2+L_1^2\right)
\NN\\&&
+\biggl\{
4 a_{Qq}^{(2), \rm{PS}} \beta_{0,Q}
-\hat{\gamma}_{qg}^{(0)} a_{gq}^{(2)}
-\frac{1}{2} \beta_{0,Q} \zeta_2 \gamma_{gq}^{(0)} \hat{\gamma}_{qg}^{(0)}
\biggr\} \left(L_1+L_2\right)
+8 \overline{a}_{Qq}^{(2), \rm{PS}} \beta_{0,Q}
-2 \hat{\gamma}_{qg}^{(0)} \overline{a}_{gq}^{(2)}
\NN\\&&
+\tilde{a}_{Qq}^{(3), \rm{PS}}~.
\label{Aqq3PSQMSren}
\end{eqnarray}

\begin{figure}[t]
\begin{center}
\begin{minipage}[c]{0.20\linewidth}
  \includegraphics[width=1\textwidth]{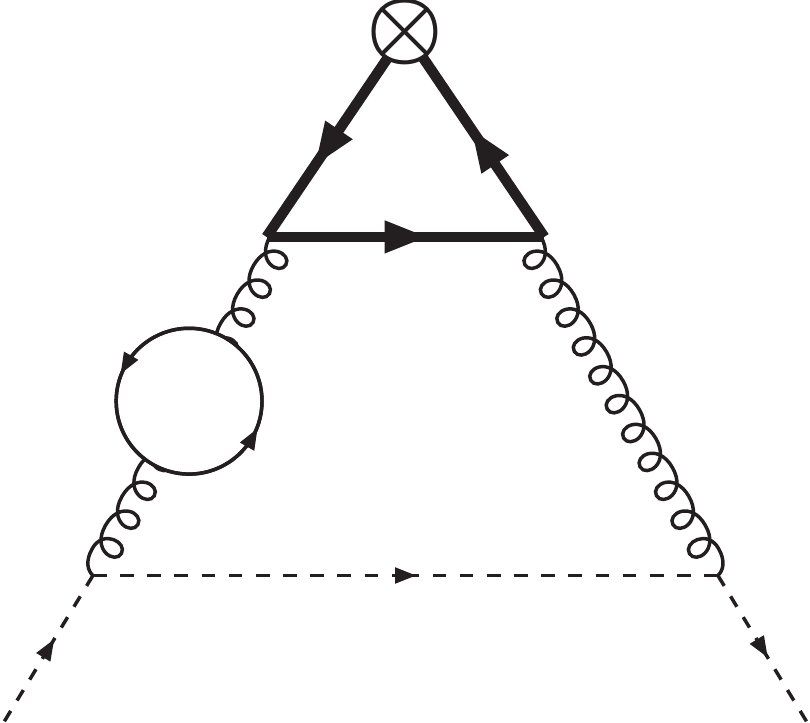}
\vspace*{-9mm}
\begin{center}
{\footnotesize (1)}
\end{center}
\end{minipage}
\hspace*{8mm}
\begin{minipage}[c]{0.20\linewidth}
  \includegraphics[width=1\textwidth]{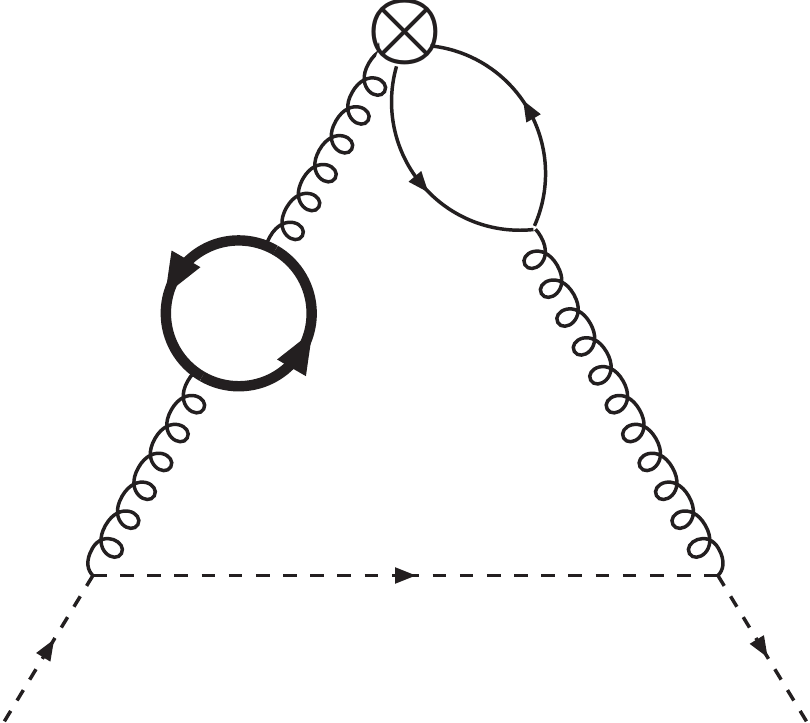}
\vspace*{-9mm}
\begin{center}
{\footnotesize (2)}
\end{center}
\end{minipage}
\hspace*{8mm}
\begin{minipage}[c]{0.20\linewidth}
  \includegraphics[width=1\textwidth]{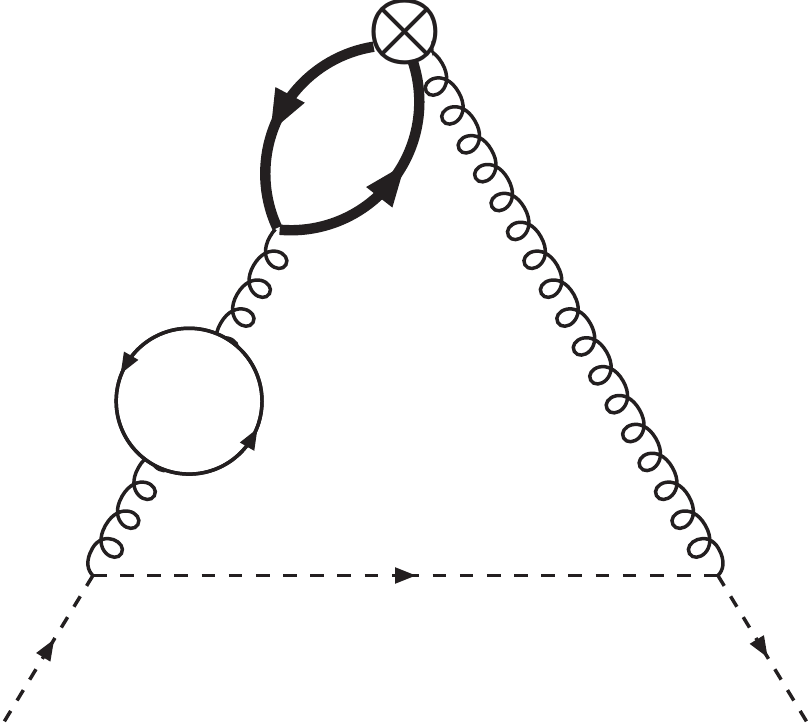}
\vspace*{-9mm}
\begin{center}
{\footnotesize (3)}
\end{center}
\end{minipage}
\caption{\small Diagrams for the two-mass contributions to $\tilde{A}_{Qq}^{(3), \rm PS}$. The curly lines denote gluons,
the dashed arrow line represents the external
massless quarks, while the thick solid arrow line represents a quark of mass $m_1$, and the thin arrow line a quark of mass $m_2$. We assume $m_1 > m_2$.}
\label{PSdiagrams}
\end{center}
\end{figure}

There are sixteen diagrams contributing in this case, three of which are shown in Figure \ref{PSdiagrams}. 
All others are related to these by the exchange of masses and/or reversal of fermion lines.

We use the same trick that we used before for $\tilde{A}_{qq}^{\rm NS}$ and $\tilde{A}_{gq}$ to decouple the mass coming 
from 
the fermion loop without the operator insertion. For the fermion loop with the vertex operator insertion, we use
\begin{eqnarray}
\raisebox{-6mm}{\includegraphics[keepaspectratio = true, scale = 0.8]{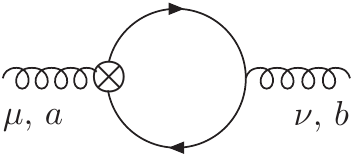}} &=&
16 \delta_{ab} T_F g_s^2 \frac{(\Delta.k)^{N-2}}{(4\pi)^{D/2}} \Gamma(2-D/2)
\int_0^1 dx \,\, x^N (1-x) \frac{(\Delta.k) \Delta_{\mu} k_{\nu} - k^2 \Delta_{\mu} \Delta_{\nu}}{(m^2-x (1-x) k^2)^{2-D/2}}.
\nonumber \\ &&
\end{eqnarray}
One obtains the following result for diagram 3 in Figure \ref{PSdiagrams},
\begin{eqnarray}
D_3(N) &=& -64 C_F T_F^2 \left(1+(-1)^N\right) \left(\frac{m_2^2}{\mu^2}\right)^{\frac{3}{2} \ep} \left(1+\frac{\ep}{2}\right)
\frac{\Gamma(N-1)}{\Gamma(N+1+\ep/2)} 
\nonumber \\ && \times
\frac{1}{2 \pi i} \int_{-i \infty}^{+i \infty} d\sigma \, \eta^{-\sigma} \,
\frac{\Gamma(-\sigma+N+1+\ep/2) \Gamma(-\sigma+2+\ep/2)}{\Gamma(-2 \sigma+N+3+\ep)}
\Gamma(-\sigma) \Gamma(-\sigma+\ep) 
\nonumber \\ && 
\phantom{\frac{1}{2 \pi i} \int_{-i \infty}^{+i \infty} d\sigma \, } \times
\Gamma\left(\sigma-\frac{3}{2} \ep\right)
\Gamma(\sigma-\ep/2) \frac{\Gamma^2(\sigma+2-\ep)}{\Gamma(2 \sigma+4-2 \ep)}.
\label{D3}
\end{eqnarray}
The contour integral in Eq. (\ref{D3}) can be performed using the Mathematica package {\tt MB}~\cite{MB}, together with the extension {\tt MBresolve}~\cite{MBr},
which allows to resolve the singularity structure in $\ep$ of this type of integrals by taking residues in $\sigma$. Once this is done,
we can expand in $\ep$. At $O(\ep^0)$, we are left with a sum of terms containing the poles in $\ep$ of the integral and the original integral 
with $\ep$ set to zero.

In the case where the operator insertion lies on a fermion line, we use the following expression,
\begin{eqnarray}
\raisebox{-6mm}{\includegraphics[keepaspectratio = true, scale = 0.8]{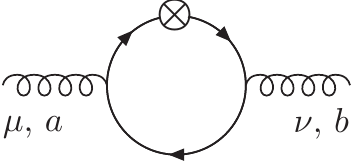}} &=&
4 \delta_{ab} T_F g_s^2 \frac{(\Delta.k)^{N-2}}{(4\pi)^{D/2}} \int_0^1 dx \,\, x^{N-2} (1-x) \biggl[ 
\nonumber \\ &&
-2 \left(x (1-x) (g_{\mu \nu} k^2 -2 k_{\mu} k_{\nu})+m^2 g_{\mu \nu}\right) \frac{x^2 \Gamma(3-D/2) (\Delta.k)^2}{(m^2-x (1-x) k^2)^{3-D/2}}
\nonumber \\ &&
+\Gamma(2-D/2) (2 N x+1-N) \frac{x (k_{\mu} \Delta_{\nu}+k_{\nu} \Delta_{\mu})(\Delta.k)}{(m^2-x (1-x) k^2)^{2-D/2}} 
\nonumber \\ &&
+\Gamma(2-D/2) ((N-1) (1-2 x)-D x) \frac{x g_{\mu \nu} (\Delta.k)^2}{(m^2-x (1-x) k^2)^{2-D/2}} 
\nonumber \\ &&
-\Gamma(1-D/2) \frac{N-1}{1-x} (N (1-x)-1) \frac{\Delta_{\mu} \Delta_{\nu}}{(m^2-x (1-x) k^2)^{1-D/2}} \biggr],
\end{eqnarray}
which allows us to write the result for diagram 1 in Figure \ref{PSdiagrams} (as well as all related diagrams) in terms
of a linear combination of contour integrals similar to the one appearing in Eq.~(\ref{D3}).

The $\Gamma$--functions in the integrand of Eq. (\ref{D3}) arise from the integration of Feynman parameters into Euler Beta--functions.
In Ref.~\cite{Ablinger:2017xml}, we left one of these Feynman parameters unintegrated, since it was already in the 
form of a
Mellin transform. In the case of Eq. (\ref{D3}) this refers to the term
\begin{eqnarray}
\frac{\Gamma(-\sigma+N+1+\ep/2) \Gamma(-\sigma+2+\ep/2)}{\Gamma(-2 \sigma+N+3+\ep)} = 
\int_0^1 dx \, x^{-\sigma+N+\ep/2} (1-x)^{-\sigma+1+\ep/2}.
\end{eqnarray}
The exchange $m_1 \leftrightarrow m_2$ implies also the change $\eta \rightarrow 1/\eta$, which means that in 
Ref.~\cite{Ablinger:2017xml},
for some of the diagrams the integrals depended on $(\eta x (1-x))^{-1}$, while for other diagrams, the dependence was on $\frac{\eta}{x (1-x)}$. 
In the former case, we can close the contour to the left and take residues, since $(\eta x (1-x))^{-1} > 4$ for all values of $x \in (0,1)$. 
The latter case is a bit more complicated, since $\frac{\eta}{x (1-x)}$ will be smaller or bigger than 4, for $\eta<1$, depending on the value of $x$, 
which means that in some regions of $x$ we had to close the contours to the left, while in the remaining regions we had to close it to the right. On top of
this, there are factors depending on $N$, such as the factor $\Gamma(N-1)/\Gamma(N+1+\ep/2)$ in (\ref{D3}), which after the $\ep$ expansion,
lead to factors of the form $1/N$ and $1/(N+1)$. These factors were absorbed into the Mellin transform using integration by parts, and the
final result was then given in terms of integrals weighted by Heaviside $\theta$--functions.

In these proceedings we show an alternative method of calculating the pure singlet diagrams. We work directly in the representation of the diagrams 
with the fully integrated Feynman parameters. Let us consider the contour integral in (\ref{D3}). After the $\ep$ 
expansion, one of the terms
we get is given by
\begin{eqnarray}
J=\frac{1}{N+1} \frac{1}{2\pi i} \int_{-1/2-i \infty}^{-1/2+i \infty} d\sigma \;\; \eta^{-\sigma}
\frac{\Gamma(N+1-\sigma) \Gamma(-\sigma)^2 \Gamma(\sigma)^2 \Gamma(2-\sigma) \Gamma(2+\sigma)^2}{\Gamma(N+3-2\sigma) \Gamma(4+2\sigma)}.
\end{eqnarray}
There is also a term with the same contour integral but a factor $1/N$ instead of $1/(N+1)$ in the decomposition of the diagram, 
and in the case of the diagrams where the operator insertion lies on a fermion line, also the factor $1/(N-1)$ appears in some cases.

Closing the contour to the left and summing residues, we obtain the following result
\begin{equation}
J = \eta \frac{\Gamma(N+1)}{\Gamma(N+4)} \big(4 S_1(N+3)-2 S_1(N)-2 \ln(\eta)-3\big)+J',
\end{equation}
where
\begin{eqnarray}
J' &=& \sum_{k=2}^{\infty} \frac{\eta^k}{N+1} B(k+N+1,k+2) \frac{ \Gamma (k)^2 \Gamma (2 k-3)}{\Gamma(k-1)^2 \Gamma(k+1)^2} \Biggl\{
\ln^2(\eta)+2 \ln(\eta) \biggl[S_1(k+N)
\nonumber \\ &&
-2 S_1(2 k+N+2)
-2 S_1(k-2)
-\frac{2}{k}
+S_1(k+1)
+2 S_1(2 k-4)\biggr]
+\frac{2}{k^2}
\nonumber \\ &&
+\biggl(S_1(k+N)
-2 S_1(2 k+N+2)
-2 S_1(k-2)
-\frac{2}{k}
+S_1(k+1)
+2 S_1(2 k-4)\biggr)^2
\nonumber \\ &&
-S_2(k+N)
+4 S_2(2 k+N+2)
+2 S_2(k-2)
-S_2(k+1)-4 S_2(2 k-4)
\Biggr\}.
\end{eqnarray}
Unfortunately, this sum, as it stands, cannot be done using {\tt Sigma}. Therefore, we proceed as follows. First we 
reintroduce an integral in $x$ using\footnote{Transformations of this kind have been used in Ref.~\cite{Bierenbaum:2007dm} 
before.}
\begin{eqnarray}
\int_0^1 dx \; x^{k+N} (1-x)^{k+1} &=& B(k+N+1,k+2) \\
\int_0^1 dx \; x^{k+N} (1-x)^{k+1} \ln(x) &=& B(k+N+1,k+2) \left(S_1(k+N)-S_1(2k+N+2)\right), \nonumber \\&&
\\
\int_0^1 dx \; x^{k+N} (1-x)^{k+1} \ln(1-x) &=& B(k+N+1,k+2) \left(S_1(k+1)-S_1(2k+N+2)\right), \nonumber \\&&
\\
\int_0^1 dx \; x^{k+N} (1-x)^{k+1} \ln(x)^2 &=& B(k+N+1,k+2) \bigl[\left(S_1(k+N)-S_1(2k+N+2)\right)^2
\nonumber \\ &&
-S_2(k+N)+S_2(2k+N+2)\bigr],
\\
\int_0^1 dx \; x^{k+N} (1-x)^{k+1} \ln(1-x)^2 &=& B(k+N+1,k+2) \bigl[\left(S_1(k+1)-S_1(2k+N+2)\right)^2
\nonumber \\ &&
-S_2(k+1)+S_2(2k+N+2)\bigr], 
\\
\int_0^1 dx \; x^{k+N} (1-x)^{k+1} \ln(x) \ln(1-x) &=& B(k+N+1,k+2) \bigl[-\zeta(2)+S_2(2k+N+2)
\nonumber \\ &&
+\left(S_1(k+1)-S_1(2k+N+2)\right)
\nonumber \\ && \times
\left(S_1(k+N)-S_1(2k+N+2)\right) \bigr],
\end{eqnarray}
which leads to
\begin{eqnarray}
J' &=& \sum_{k=2}^{\infty} \frac{\eta^k}{N+1} \int_0^1 dx \; (1-x)^{k+1} x^{k+N}
\frac{\Gamma (k)^2 \Gamma (2 k-3)}{\Gamma(k-1)^2 \Gamma(k+1)^2} \biggl\{\ln^2(\eta)
+2 \ln(\eta) \biggl[-2 S_1(k-2)
\nonumber \\ &&
-\frac{2}{k}
+2 S_1(2 k-4)
+\ln(1-x)+\ln(x)\biggr]
-4 \biggl[S_1(k-2)
+\frac{1}{k}
-S_1(2 k-4)\biggr] (\ln(1-x)+\ln(x))
\nonumber \\ &&
+4 \biggl(S_1(k-2)+\frac{1}{k}-S_1(2 k-4)\biggr)^2
+2 \zeta_2
+2 S_2(k-2)
+\frac{2}{k^2}
-4 S_2(2 k-4)
\nonumber \\ &&
+\left(\ln(1-x)+\ln(x)\right)^2
\biggr\}.
\end{eqnarray}
Now we do the binomial expansion of the term $(1-x)^{k+1}$ and use
\begin{equation}
\frac{1}{N+l} \int_0^1 dx \; x^N f(x) = \int_0^1 dx \; x^{N+l-1} \left(\int_0^1 dy \; y^{-l} f(y) - \int_0^x dy \; y^{-l} f(y)\right),
\end{equation}
in order to absorb the $\frac{1}{N+l}$ factor into the integrand. The integrals from 0 to $x$ can be done using
\begin{eqnarray}
\int_0^x dy \; y^n \ln(y) &=& \frac{x^{n+1}}{n+1} \ln(x)-\frac{x^{n+1}}{(n+1)^2}, \\
\int_0^x dy \; y^n \ln(1-y) &=& -\frac{S_1(x,n)}{n+1}-\frac{x^{n+1}}{(n+1)^2}+\frac{x^{n+1}-1}{n+1} \ln(1-x), \\
\int_0^x dy \; y^n \ln(y)^2 &=& \frac{2 x^{n+1}}{(n+1)^3}+\frac{x^{n+1}}{n+1} \ln^2(x)-\frac{2 x^{n+1}}{(n+1)^2} \ln(x), \\
\int_0^x dy \; y^n \ln(1-y)^2 &=& \frac{\ln^2(x)}{n+1} \biggl[\frac{2}{n+1} S_1(x,n+1)+2 S_{1,1}(\{1,x\},n)-2 x
\nonumber \\&&
+2 \ln(1-x) \big(S_1(n+1)-S_1(x,n+1)-1+x\big)
+2 x
\nonumber \\&&
-x \left(1-x^n\right) \ln^2(1-x)
+(1-x) \big(2 \ln(1-x)-\ln^2(1-x)\big)\biggr], 
\\
\int_0^x dy \; y^n \ln(y) \ln(1-y) &=& 
\frac{1}{(n+1)^2} \Big[S_1(x,n-1)-x^{n+1} \ln(x)-\left(x^{n+1}-1\right) \ln(1-x)\Big]
\nonumber \\&&
-\frac{x^n \ln(x)}{n (n+1)}
+\frac{1}{n+1} \Big[S_2(x,n-1)-\ln(x) S_1(x,n-1)-\text{Li}_2(x)
\nonumber \\&&
+\left(x^{n+1}-1\right) \ln(x) \ln(1-x)\Big]
+\frac{(2 n+1) x^n}{n^2 (n+1)^2}
+\frac{2 x^{n+1}}{(n+1)^3}.
\end{eqnarray}
From this we get,
\begin{eqnarray}
J' &=& \int_0^1 dx \,\, x^{N+1} \sum_{k=2}^{\infty} \sum_{i=0}^{k+1} (-1)^i \eta^k \binom{k+1}{i}
\frac{\Gamma (k)^2 \Gamma (2 k-3)}{\Gamma (k-1)^2 \Gamma(k+1)^2}
\Biggl\{
\frac{8 \left(1-x^{i+k-1}\right)}{(i+k-1)^3}
\nonumber \\ &&
+\frac{1}{(i+k-1)^2} \biggl[-4 S_1(x,i+k-2)+4 S_1(i+k-2)+\left(1-x^{i+k-1}\right) \biggl(8 S_1(k-2)
\nonumber \\ &&
+\frac{8}{k}
-8 S_1(2 k-4)-4 \ln(\eta )\biggr)
+4 x^{i+k-1} \ln(x) 
+2 \left(2 x^{i+k-1}-1\right) \ln(1-x) \biggr]
\nonumber \\ &&
+\frac{1}{i+k-1} \biggl\{\ln(\eta) \biggl(2 S_1(x,i+k-2)
-2 S_1(i+k-2)
-2 x^{i+k-1} \ln(x)\biggr)
\nonumber \\ &&
+\ln(x) \biggl(2 S_1(x,i+k-2)+4 x^{i+k-1} S_1(k-2) 
+\frac{4}{k} x^{i+k-1}
-4 x^{i+k-1} S_1(2 k-4)\biggr)
\nonumber \\ &&
+4 \biggl(S_1(k-2)+\frac{1}{k}-S_1(2 k-4)\biggr) \big(S_1(i+k-2)-S_1(x,i+k-2)\big)
+2 \text{Li}_2(x)
\nonumber \\ &&
-2 S_2(x,i+k-2)-2 S_{1,1}(\{1,x\},i+k-2)
-\frac{\pi ^2}{3} x^{i+k-1}
-x^{i+k-1} \ln^2(x) 
\nonumber \\ &&
+\ln(1-x) \big(2 S_1(x,i+k-2)-2 S_1(i+k-1)\big)+2 S_2(i+k-2)+2 S_{1,1}(i+k-2)
\nonumber \\ &&
+\left(1-x^{i+k-1}\right)
\biggl[\ln(\eta ) \biggl(-4 S_1(k-2)
-\frac{4}{k}
+4 S_1(2 k-4)+2 \ln(1-x)\biggr)
\nonumber \\ &&
-4 \ln(1-x) \biggl(S_1(k-2)
+\frac{1}{k}
-S_1(2 k-4)\biggr)
+4 \biggl(S_1(k-2)+\frac{1}{k}-S_1(2 k-4)\biggr)^2
\nonumber \\ &&
+2 S_2(k-2)
+\frac{2}{k^2}
-4 S_2(2 k-4)+\ln^2(\eta )
+\ln^2(1-x)+2 \ln(x) \ln(1-x)\biggr]
\biggr\}
\Biggr\}.
\end{eqnarray}
Now we can do perform the double sum using {\tt Sigma} and {\tt HarmonicSums}. We finally obtain,
\begin{eqnarray}
J' &=& \int_0^1 dx \, x^{N+1} \Biggl\{
-\frac{1}{24} G^2\biggl(\left\{\frac{1}{\tau},\frac{\sqrt{4 \tau -1}}{\tau }\right\},x \eta -x^2 \eta \biggr)
-\frac{1}{12} i G\biggl(\left\{\frac{\sqrt{4 \tau -1}}{\tau },\frac{1}{\tau },\frac{1}{\tau },\frac{1}{\tau }\right\},x-x^2\biggr)
\nonumber \\ &&
+\frac{1}{288} \ln^4((1-x) x \eta)
+\frac{12 x^2-x+6}{216 x} \ln^3((1-x) x \eta)
+\biggl[\frac{\zeta_2}{24}+\frac{66 x^2-15 \eta x-37 x+24}{108 x}
\nonumber \\ &&
-\frac{11}{36} \ln(1-\eta)\biggr] \ln^2((1-x) x \eta)
+\biggl[\frac{T_2}{54 x}+\ln(1-x) \biggl(\frac{5 \eta}{36}
+\frac{11}{36} \ln(1-\eta)\biggr)
\nonumber \\ &&
+\ln(x) \biggl(\frac{11}{12} \ln(1-\eta)-\frac{7 \eta }{12}\biggr)-\frac{11}{9} x \ln(1-\eta)\biggr] \ln((1-x) x \eta)
+\frac{\zeta_2 T_3}{27 x}+\frac{\eta T_1}{810}-\zeta_2 \eta 
\nonumber \\ &&
+\biggl\{\frac{11}{36} G\biggl(\left\{\frac{\sqrt{4 \tau -1} \sqrt{4 \eta  \tau -1}}{\tau }\right\},x-x^2\biggr)
-\frac{20}{9} \eta G\biggl(\left\{\sqrt{4 \tau -1} \sqrt{4 \eta  \tau -1}\right\},x-x^2\biggr)\biggr\} \biggl[\zeta_2
\nonumber \\ &&
+i G\biggl(\left\{\frac{1}{\tau },\frac{\sqrt{4 \tau-1}}{\tau }\right\},x \eta -x^2 \eta \biggr)\biggr]
+\frac{20}{9} i \eta G\biggl(\left\{\frac{1}{\tau },\frac{\sqrt{4 \eta  \tau -1}}{\tau },\sqrt{4 \tau -1} \sqrt{4 \eta  \tau -1}\right\},x-x^2\biggr)
\nonumber \\ &&
+\frac{1}{12} i G\biggl(\left\{\frac{1}{\tau },\frac{\sqrt{4 \tau -1}}{\tau }\right\},x \eta -x^2 \eta \biggr)
\biggl[\zeta_2+G\biggl(\left\{\frac{\sqrt{4 \tau -1}}{\tau },\frac{\sqrt{4 \eta  \tau -1}}{\tau }\right\},x-x^2\biggr)\biggr]
\nonumber \\ &&
-\frac{11}{36} i G\biggl(\left\{\frac{1}{\tau },\frac{\sqrt{4 \eta  \tau -1}}{\tau },\frac{\sqrt{4 \tau -1} \sqrt{4 \eta  \tau -1}}{\tau }\right\},x-x^2\biggr)
+\frac{T_3}{27 x} \big(1-4 x (1-x) \eta\big)^{3/2} \biggl[-\zeta_2
\nonumber \\ &&
+2 i \biggl(1+\frac{1}{2} \ln((1-x) x \eta)\biggr) G\biggl(\left\{\frac{\sqrt{4 \tau -1}}{\tau }\right\},x \eta -x^2 \eta \biggr)
-i G\biggl(\left\{\frac{1}{\tau },\frac{\sqrt{4 \tau -1}}{\tau }\right\},x \eta -x^2 \eta \biggr)
\biggr]
\nonumber \\ &&
+\frac{1}{12} i G\biggl(\left\{\frac{1}{\tau },\frac{\sqrt{4 \eta  \tau -1}}{\tau },\frac{\sqrt{4 \eta  \tau -1}}{\tau },\frac{\sqrt{4 \tau -1}}{\tau}\right\},x-x^2\biggr)
+\ln^2(x) \biggl(\frac{7 \eta }{24}-\frac{11}{24} \ln(1-\eta)\biggr)
\nonumber \\ &&
+\ln^2(1-x) \biggl(-\frac{5 \eta }{72}-\frac{11}{72} \ln(1-\eta)\biggr)
+\ln(1-x) \biggl(\frac{11}{18} \ln(1-\eta)-\frac{1}{27} \eta  (8 \eta -29)\biggr)
\nonumber \\ &&
+\ln(x) \biggl[-\frac{1}{54} \eta  (16 \eta -169)+\ln(1-x) \biggl(\frac{7 \eta }{12}-\frac{11}{12} \ln(1-\eta)\biggr)+\frac{11}{18} \ln(1-\eta)\biggr]
\nonumber \\ &&
-\frac{i}{12} \biggl(\ln(\eta)+\frac{17}{3}\biggr) G\biggl(\left\{\frac{\sqrt{4 \tau -1}}{\tau },\frac{1}{\tau },\frac{1}{\tau }\right\},x-x^2\biggr)
+\biggl(\frac{\ln(x)}{6}
-\frac{\ln(\eta)}{3}\biggr)
\text{Li}_3(\eta)
+\frac{\text{Li}_4(\eta)}{2}
\nonumber \\ &&
+\biggl(\frac{1}{12} \ln((1-x) x \eta)-\frac{12 x^2-13 x+6}{36 x}\biggr) G\biggl(\left\{\frac{1}{\tau },\frac{\sqrt{4 \tau -1}}{\tau },\frac{\sqrt{4 \tau -1}}{\tau }\right\},x \eta -x^2 \eta \biggr)
\nonumber \\ &&
+\biggl(1+\frac{1}{2} \ln((1-x) x \eta)\biggr) \biggl\{
-\frac{1}{6} i G\biggl(\left\{\frac{\sqrt{4 \tau -1}}{\tau},\frac{\sqrt{4 \eta  \tau -1}}{\tau },\frac{\sqrt{4 \eta  \tau -1}}{\tau }\right\},x-x^2\biggr)
\nonumber \\ &&
+\frac{T_4}{36 x} G^2\biggl(\left\{\frac{\sqrt{4 \tau -1}}{\tau }\right\},x \eta -x^2\eta \biggr)
+\biggl[\frac{40}{9} i \eta  G\biggl(\left\{\sqrt{4 \tau -1} \sqrt{4 \eta  \tau -1}\right\},x-x^2\biggr)
\nonumber \\ &&
-\frac{11}{18} i G\biggl(\left\{\frac{\sqrt{4 \tau -1} \sqrt{4 \eta  \tau -1}}{\tau }\right\},x-x^2\biggr)\biggr] G\biggl(\left\{\frac{\sqrt{4 \tau -1}}{\tau }\right\},x \eta-x^2 \eta \biggr)
\biggr\}
\nonumber \\ &&
-i \eta \biggl(\frac{40}{9}+\frac{20}{9} \ln((1-x) x \eta)\biggr) G\biggl(\left\{\frac{\sqrt{4 \eta  \tau -1}}{\tau },\sqrt{4 \tau -1} \sqrt{4 \eta  \tau -1}\right\},x-x^2\biggr)
\nonumber \\ &&
+i\biggl(\frac{11}{18}+\frac{11}{36} \ln((1-x) x \eta)\biggr)  G\biggl(\left\{\frac{\sqrt{4 \eta  \tau -1}}{\tau },\frac{\sqrt{4 \tau -1} \sqrt{4 \eta  \tau -1}}{\tau }\right\},x-x^2\biggr)
\nonumber \\ &&
+\frac{T_4}{36 x} \biggl[i \zeta_2 G\biggl(\left\{\frac{\sqrt{4 \tau -1}}{\tau }\right\},x \eta -x^2 \eta \biggr)
-G\biggl(\left\{\frac{\sqrt{4 \tau-1}}{\tau },\frac{1}{\tau },\frac{\sqrt{4 \tau -1}}{\tau }\right\},x \eta -x^2 \eta \biggr)
\nonumber \\ &&
+\zeta_2 \ln((1-x) x \eta)\biggr]
+\text{Li}_2(x) \biggl(\frac{13 \eta }{18}-\frac{11}{18} \ln(1-\eta)+\frac{\text{Li}_2(\eta)}{6}\biggr)
-\frac{1}{6} \ln(1-x) \text{Li}_3(\eta)
\nonumber \\ &&
+G\biggl(\left\{\frac{\sqrt{4 \tau -1}}{\tau },\frac{\sqrt{4 \eta  \tau -1}}{\tau }\right\},x-x^2\biggr) \biggl(
\frac{\zeta_2}{12}+\frac{\ln(\eta)}{6}+\frac{1}{12} \ln(\eta) \ln((1-x) x \eta)\biggr)
\nonumber \\ &&
+\frac{i}{12} G\biggl(\left\{\frac{1}{\tau },\frac{\sqrt{4 \tau -1}}{\tau }\right\},x-x^2\biggr) \biggl[
\frac{1}{2} \ln^2(\eta)
+\frac{17}{3} \ln(\eta)
+\zeta_2
-\frac{1}{3} (5 \eta -22)
-\frac{11}{3} \ln(1-\eta)
\nonumber \\ &&
+\text{Li}_2(\eta)\biggr]
+\biggl[\frac{1}{24} \ln^2(1-x)+\biggl(\frac{1}{36}
(12 x-17)+\frac{1}{12} \ln(x \eta)\biggr) \ln(1-x)+\frac{\ln^2(\eta)}{8}-\frac{1}{24} \ln^2(x \eta)
\nonumber \\ &&
+\frac{1}{18} x \left(1+\sqrt{\eta}\right) \left(5 \eta -5 \sqrt{\eta }-22\right)
+\frac{1}{36} (12 x+5) \ln(x)+\frac{1}{18} (6 x-11) \ln(\eta)\biggr] \text{Li}_2(\eta)
\nonumber \\ &&
+\sqrt{\eta} \left(5 \eta-27\right) \biggl\{
\ln\biggl(\frac{1+\sqrt{\eta}}{1-\sqrt{\eta }}\biggr) \biggl[\frac{1}{144} \ln^2(1-x)+\biggl(\frac{1}{36} (2 x-1)+\frac{1}{72} \ln(x \eta)\biggr) \ln(1-x)
\nonumber \\ &&
+\frac{\ln^2(\eta)}{48}-\frac{1}{144} \ln^2(x \eta)+\frac{1}{72} i G\biggl(\left\{\frac{1}{\tau },\frac{\sqrt{4 \tau -1}}{\tau }\right\},x-x^2\biggr)+\frac{1}{36} (2 x-1) \ln(x)+\frac{x}{18} \ln(\eta)
\nonumber \\ &&
+\frac{\text{Li}_2(x)}{36}\biggr]
+\biggl(-\frac{2 x}{9}-\frac{1}{18} \ln(1-x)+\frac{\ln(x)}{18}-\frac{\ln(\eta)}{9}\biggr)
\text{Li}_2\left(\sqrt{\eta }\right)
+\frac{2}{9} \text{Li}_3\left(\sqrt{\eta}\right)-\frac{\text{Li}_3(\eta)}{36}
\nonumber \\ &&
-\frac{i}{18} G\biggl(\left\{\frac{\sqrt{4 \tau -1}}{\tau }\right\},x-x^2\biggr) \biggl[
\nonumber \\ &&
\frac{1}{2} \ln\biggl(\frac{1+\sqrt{\eta}}{1-\sqrt{\eta }}\biggr) \biggl(1+\frac{1}{2} \ln((1-x) x \eta)\biggr)
-\text{Li}_2\left(\sqrt{\eta }\right)
\nonumber \\ &&
+\frac{\text{Li}_2(\eta)}{4}\biggr]
+\biggl(\frac{1}{72} \ln(1-x)-\frac{\ln(x)}{72}+\frac{\ln(\eta)}{36}\biggr) \text{Li}_2(\eta)
\biggr\}
+\biggl[
\frac{1}{36} i \ln^3(\eta)-\frac{11}{72} i \ln^2(\eta)
\nonumber \\ &&
+\frac{1}{36} i (5 \eta -22) \ln(\eta)
-\frac{1}{12} G\biggl(\left\{\frac{1}{\tau },\frac{\sqrt{4 \tau -1}}{\tau }\right\},x \eta-x^2 \eta \biggr) \ln(\eta)
-\frac{11 i \zeta_2}{36}
-\frac{1}{27} i \eta  (8 \eta -29)
\nonumber \\ &&
-\frac{1}{12} i G\biggl(\left\{\frac{1}{\tau },\frac{\sqrt{4 \tau-1}}{\tau },\frac{\sqrt{4 \tau -1}}{\tau }\right\},x \eta -x^2 \eta \biggr)
+\frac{i}{36} \biggl(
5 \eta-3 \zeta_2-22
+11 \ln(1-\eta)
\nonumber \\ &&
-17 \ln(\eta)\biggr) \ln((1-x) x)
+\frac{11}{18} i \ln(1-\eta) \biggl(1+\frac{\ln(\eta)}{2}\biggr)
+\biggl(\frac{5 i}{36}
-\frac{1}{12} i \ln((1-x) x \eta)\biggr)
\text{Li}_2(\eta)
\nonumber \\ &&
+\frac{i \text{Li}_3(\eta)}{6}\biggr]
G\biggl(\left\{\frac{\sqrt{4 \tau -1}}{\tau }\right\},x-x^2\biggr)
+\frac{1}{18} (11-12 x) \text{Li}_3(\eta)
\Biggr\},
\end{eqnarray}
where
\begin{eqnarray}
T_1 &=& 5 \left(640 x^3-1011 x^2-171 x+810\right)+32 \eta  x \left(144 x^4-360 x^3+320 x^2-120 x+15\right), \\
T_2 &=& 192 \eta ^2 x^6-512 \eta ^2 x^5+32 \eta  (18 \eta +5) x^4
-\eta  (416 \eta +251) x^3
\nonumber \\ &&
+2 \left(96 \eta ^2+103 \eta +30\right) x^2 
-(231 \eta+32) x+12, \\
T_3 &=& 15 x^2-8 x+3, \\
T_4 &=& 12 x^2-7 x+6.
\end{eqnarray}
All other integrals appearing in the calculation can be done in a similar way. This way of calculating the diagrams applies perfectly well in all of the cases
where the operator insertion lies on the heaviest of the quarks, as long as we are assuming that $\eta<1$. Taking into account that
the diagrams where the insertion lies on the lightest quark are related to the former diagrams by the change $\eta \rightarrow 1/\eta$, we
can in principle obtain the results of the latter diagrams by analytic continuation to the region where $\eta>1$. This is, however, far less trivial
compared with the $\tilde{A}_{gg,Q}$ case, since the square root letters in the alphabet of the iterated integrals introduce branch cuts that need to be
analyzed carefully. The complete result given in Ref.~\cite{Ablinger:2017xml} has been obtained in a different way.

\section{Conclusions \label{CONC}}

\vspace*{1mm}
\noindent
We have presented the two--mass contributions for a series of three--loop OMEs. The simplest ones are $\tilde{A}_{qq,Q}^{(3), \rm NS}$ and
$\tilde{A}_{gq,Q}^{(3)}$, for which the $N$ and $\eta$ dependence factorizes. In the case of $\tilde{A}_{gg,Q}^{(3)}$, 
we presented results in the physical case, going beyond those given in \cite{Ablinger:2017err}, where a series of scalar diagrams were computed.
In the pure singlet case, which was studied in \cite{Ablinger:2017xml}, we presented a new way of computing the diagrams. This new way has the advantage of producing
a result that is given entirely by iterated integrals, without the need to introduce additional integrations on top of these, as we did 
in \cite{Ablinger:2017xml}. However, the results so far are only valid for half of the diagrams. In order to cover the other half, the analytic continuation 
$\eta \rightarrow 1/\eta$ still remain to be done. 

With the full results for $\tilde{A}_{gg,Q}^{(3)}$ within grasp, and having calculated all the other OMEs presented here, the only remaining OME is 
$\tilde{A}_{Qg}^{(3)}$. Considering that even in the single mass case this OME exhibits integrals with an elliptic 
behaviour~\cite{Ablinger:2017bjx,Ablinger:2017cin}, it doesn't seem feasible that this OME will be computed analytically
any time soon. New methods will need to be applied in this case, which we plan to study at some point in the future.



\begin{thebibliography}{99}
%
\bibitem{ALPHA}
  S.~Bethke {\it  et al.}, 
  {\it  Workshop on Precision Measurements of $\alpha_s$},
  arXiv:1110.0016 [hep-ph];\\
  S.~Moch, S.~Weinzierl {\it  et al.}, 
  {\it  High precision fundamental constants at the TeV scale},
  arXiv:1405.4781 [hep-ph];\\
  S.~Alekhin, J.~Bl\"umlein and S.O.~Moch,
  Mod.\ Phys.\ Lett.\ A {\bf 31} (2016) no.25,  1630023.
%
\bibitem{PDF}
  A.~Accardi {\it  et al.},
  Eur.\ Phys.\ J.\ C {\bf 76} (2016) no.8,  471
  [arXiv:1603.08906 [hep-ph]];\\
  S.~Alekhin, J.~Bl\"umlein, S.~Moch and R.~Placakyte,
  Phys.\ Rev.\ D {\bf 96} (2017) no.1,  014011
  [arXiv:1701.05838 [hep-ph]].
%
\bibitem{MCMB}
  S.~Alekhin, J.~Bl\"umlein, K.~Daum, K.~Lipka and S.~Moch,
  Phys.\ Lett.\ B {\bf 720} (2013) 172
  [arXiv:1212.2355 [hep-ph]];\\
  A.~Gizhko {\it et al.},
  Phys.\ Lett.\ B {\bf 775} (2017) 233
  [arXiv:1705.08863 [hep-ph]].
%
\bibitem{Ablinger:2014lka}
  J.~Ablinger, J.~Bl\"umlein, A.~De Freitas, A.~Hasselhuhn, A.~von Manteuffel, M.~Round, C.~Schneider and F.~Wi{\ss}brock,
  Nucl.\ Phys.\ B {\bf 882} (2014) 263
  [arXiv:1402.0359 [hep-ph]].
%
\bibitem{Ablinger:2014uka}
  J.~Ablinger, J.~Bl\"umlein, A.~De Freitas, A.~Hasselhuhn, A.~von Manteuffel, M.~Round and C.~Schneider,
  Nucl.\ Phys.\ B {\bf 885} (2014) 280
  [arXiv:1405.4259 [hep-ph]].
%
\bibitem{Ablinger:2014vwa}
J.~Ablinger, A.~Behring, J.~Bl\"umlein, A.~De Freitas, A.~Hasselhuhn, A.~von Manteuffel, 
M.~Round, C.~Schneider, F.~Wi\ss{}brock,
  Nucl.\ Phys.\ B {\bf 886} (2014) 733
  [arXiv:1406.4654 [hep-ph]].
%
\bibitem{Behring:2015zaa}
  A.~Behring, J.~Bl\"umlein, A.~De Freitas, A.~von Manteuffel and C.~Schneider,
  Nucl.\ Phys.\ B {\bf 897} (2015) 612
  [arXiv:1504.08217 [hep-ph]].
%
\bibitem{Behring:2015roa}
  A.~Behring, J.~Bl\"umlein, A.~De Freitas, A.~Hasselhuhn, A.~von Manteuffel and C.~Schneider,
  Phys.\ Rev.\ D {\bf 92} (2015) no.11,  114005
  [arXiv:1508.01449 [hep-ph]].
%
\bibitem{Behring:2016hpa}
  A.~Behring, J.~Bl\"umlein, G.~Falcioni, A.~De Freitas, A.~von Manteuffel and C.~Schneider,
  Phys.\ Rev.\ D {\bf 94} (2016) no.11,  114006
  [arXiv:1609.06255 [hep-ph]].
%
\bibitem{Ablinger:2014nga}
  J.~Ablinger, A.~Behring, J.~Bl\"umlein, A.~De Freitas, A.~von Manteuffel and C.~Schneider,
  Nucl.\ Phys.\ B {\bf 890} (2014) 48
  [arXiv:1409.1135 [hep-ph]].
%
\bibitem{Ablinger:2015tua}
  J.~Ablinger, A.~Behring, J.~Bl\"umlein, A.~De Freitas, A.~von Manteuffel and C.~Schneider,
  Comput.\ Phys.\ Commun.\  {\bf 202} (2016) 33
  [arXiv:1509.08324 [hep-ph]].
%
\bibitem{Ablinger:2017ptf}
  J.~Ablinger, A.~Behring, J.~Bl\"umlein, A.~De Freitas, A.~von Manteuffel and C.~Schneider,
  PoS (QCDEV2017) 031  [arXiv:1711.07957 [hep-ph]].
%
\bibitem{Blumlein:2017dxp}
  J.~Bl\"umlein and C.~Schneider,
  Phys.\ Lett.\ B {\bf 771} (2017) 31
  [arXiv:1701.04614 [hep-ph]].
%
\bibitem{Behring:2014eya}
  A.~Behring, I.~Bierenbaum, J.~Bl\"umlein, A.~De Freitas, S.~Klein and F.~Wi{\ss}brock,
  Eur.\ Phys.\ J.\ C {\bf 74} (2014) no.9,  3033
  [arXiv:1403.6356 [hep-ph]].
%
\bibitem{Bierenbaum:2009mv}
  I.~Bierenbaum, J.~Bl\"umlein and S.~Klein,
  Nucl.\ Phys.\ B {\bf 820} (2009) 417
  [arXiv:0904.3563 [hep-ph]].
%
\bibitem{VFNS}
J.~Bl\"umlein, A.~De Freitas, C.~Schneider, and K.~Sch\"onwald, DESY 17-187.
%
\bibitem{Ablinger:2017err}
  J.~Ablinger, J.~Bl\"umlein, A.~De Freitas, A.~Hasselhuhn, C.~Schneider and F.~Wi{\ss}brock,
  Nucl.\ Phys.\ B {\bf 921} (2017) 585
  [arXiv:1705.07030 [hep-ph]].
%
\bibitem{Ablinger:2017xml}
  J.~Ablinger, J.~Bl\"umlein, A.~De Freitas, C.~Schneider and K.~Sch\"onwald,
  {\it The two-mass contribution to the three-loop pure singlet operator matrix element},
  arXiv:1711.06717 [hep-ph].
%
\bibitem{MB1a}
E.W.~Barnes, 
Proc. Lond. Math. Soc. (2) {\bf 6} (1908) 141.
%
\bibitem{MB1b}
E.W.~Barnes,
Quarterly Journal of Mathematics {\bf 41} (1910) 136.
%
\bibitem{MB2}
H.~Mellin,
Math. Ann. {\bf 68}, no. 3 (1910) 305.
%
\bibitem{MB3}
E.T.~Whittaker and G.N.~Watson, {\it A Course of Modern Analysis}, (Cambridge University Press, Cambridge, 1927;
 reprinted 1996) 616~p.
%
\bibitem{MB4}
E.C.~Titchmarsh,
{\it Introduction to the Theory of Fourier Integrals},
(Calendron Press, Oxford, 1937; 2nd Edition 1948).
%
\bibitem{Ablinger:2010ty}
  J.~Ablinger, J.~Bl\"umlein, S.~Klein, C.~Schneider and F.~Wi\ss{}brock,
  Nucl.\ Phys.\ B {\bf 844} (2011) 26
  [arXiv:1008.3347 [hep-ph]].
%
\bibitem{Blumlein:2012vq}
  J.~Bl\"umlein, A.~Hasselhuhn, S.~Klein and C.~Schneider,
  Nucl.\ Phys.\ B {\bf 866} (2013) 196
  [arXiv:1205.4184 [hep-ph]].
%
\bibitem{HYP}
W.N.~Bailey, {\it Generalized Hypergeometric Series}, (Cambridge University
Press,  Cambridge, 1935);\\
L.J.~Slater, {\it Generalized Hypergeometric Functions}, (Cambridge University
Press, Cambridge, 1966);\\
P.~Appell and J.~Kamp\'{e} de F\'{e}riet, {\it Fonctions
Hyperg\'{e}om\'{e}triques et Hypersp\'{e}riques, Polynomes D' Hermite},
(Gauthier-Villars, Paris, 1926);\\
P.~Appell, {\it Les Fonctions Hyperg\"{e}om\'{e}triques de Plusieur
Variables}, (Gauthier-Villars, Paris, 1925);\\
J.~Kamp\'{e} de F\'{e}riet, {\it La fonction
hyperg\"{e}om\'{e}trique},(Gauthier-Villars, Paris, 1937);\\
H.~Exton, {\it Multiple Hypergeometric Functions and Applications},
(Ellis Horwood, Chichester, 1976);\\
H.~Exton, {\it Handbook of Hypergeometric Integrals},
(Ellis Horwood, Chichester, 1978);\\
H.M.~Srivastava and P.W. Karlsson, {\it Multiple Gaussian Hypergeometric
Series}, (Ellis Horwood, Chicester, 1985);\\
  M.J.~Schlosser, in: {\it Computer Algebra in Quantum Field Theory: Integration, Summation and
  Special Functions}, C. Schneider, J. Bl\"umlein, Eds.,~p.~305, (Springer, Wien, 2013)
  [arXiv:1305.1966 [math.CA]].
%
\bibitem{Karr:81}
M.~Karr,
{J.~ACM} {\bf 28} (1981) 305.
%
\bibitem{Schneider:01}
C.~Schneider,
{\it Symbolic Summation in Difference Fields\/} Ph.D. Thesis
RISC, Johannes Kepler University, Linz technical report 01-17 (2001).
%
\bibitem{Schneider:05a}
C. Schneider,
An. Univ. Timisoara Ser. Mat.-Inform. {\bf 42} (2004) 163;\\
{J. Differ. Equations Appl.\/} {\bf 11} (2005) 799
;\\
Appl. Algebra Engrg. Comm. Comput. {\bf 16} (2005) 1.
%
\bibitem{Schneider:07d}
C.~Schneider,
{J. Algebra Appl.\/} {\bf 6} (2007) 415.
%
\bibitem{Schneider:10b}
C.~Schneider, {\it {Motives, Quantum Field Theory, and Pseudodifferential
  Operators}\/} ({\it Clay Mathematics Proceedings\/} Vol.~{\bf{12}} ed. A.~Carey,
  D.~Ellwood, S.~Paycha and S.~Rosenberg,(Amer. Math. Soc) (2010), 285 
  [arXiv:0904.2323].
%
\bibitem{Schneider:10c}
C.~Schneider,
{Ann. Comb.\/} {\bf 14} (2010) 533 
[arXiv:0808.2596].
%
\bibitem{Schneider:15a}
C.~Schneider,
in: Computer Algebra and Polynomials, Applications of Algebra and Number Theory, J.~Gutierrez, J.~Schicho, M.~Weimann 
(ed.), Lecture Notes in Computer Science (LNCS) 8942 (2015), 157
[arXiv:13077887 [cs.SC]].

\bibitem{Schneider:08c}
C. Schneider,
J. Symbolic Comput. {\bf 43} (2008) 611
[arXiv:0808.2543v1];
J. Symb. Comput. {\bf 72} (2016) 82
[arXiv:1408.2776 [cs.SC]];
J. Symb. Comput. 80 (2017), 616  [arXiv:1603.04285 [cs.SC]].
%
\bibitem{DFTheory}
C.~Schneider,
Ann. Comb. \textbf{9}(1) (2005) 75; 
\\   
S.A. Abramov and M.~Petkov{\v{s}}ek,
{J. Symbolic Comput.}, {\bf 45}(6) (2010) 684; 
\\
C.~Schneider,
{Appl. Algebra Engrg. Comm. Comput.}, {\bf 21}(1) (2010) 1;
\\   
C.~Schneider,
In: Symbolic and Numeric Algorithms for Scientific Computing (SYNASC), 2014, 15th International
Symposium, F.~Winkler, V.~Negru, T.~Ida, T.~Jebelean, D.~Petcu, S.~Watt, D.~Zaharie (ed.),
(2015)~pp.~26;
IEEE Computer Society, arXiv:1412.2782v1 [cs.SC].
%
\bibitem{SIG1}
C.~Schneider, {S\'em.~Lothar. Combin.\/} {\bf 56} (2007) 1, 
 article B56b.
%
\bibitem{SIG2}
C.~Schneider, {{\it Computer Algebra in Quantum Field Theory: Integration,
  Summation and Special Functions}\/} Texts and Monographs in Symbolic
  Computation eds. C.~Schneider and J.~Bl\"umlein  (Springer, Wien, 2013) 325, 
  arXiv:1304.4134 [cs.SC].
%
\bibitem{Huber:2007dx}
  T.~Huber and D.~Maitre,
  Comput.\ Phys.\ Commun.\  {\bf 178} (2008) 755
  [arXiv:0708.2443 [hep-ph]].
%
\bibitem{POLYLOG1}
L.~Lewin, {\it Dilogarithms and associated functions}, (Macdonald, London, 1958).
%
\bibitem{POLYLOG2}
L.~Lewin, {\it Polylogarithms and associated functions},
(North Holland, New York, 1981).
%
\bibitem{Devoto:1983tc}
  A.~Devoto and D.W.~Duke,
  Riv.\ Nuovo Cim.\  {\bf 7N6} (1984) 1.
%
\bibitem{Ablinger:2013jta}
  J.~Ablinger and J.~Bl\"umlein,
  {\it Harmonic Sums, Polylogarithms, Special Numbers, and their Generalizations},
  arXiv:1304.7071 [math-ph], in~: {\it Integration, Summation and Special Functions in Quantum Field 
  Theory}, 
  eds.~J.~Bl\"umlein and C.~Schneider, (Springer, Wien, 2013) 1. 
%
\bibitem{HARMONICSUMS}
  J.~Ablinger,
  PoS {(LL2014)} 019;
  {\it Computer Algebra Algorithms for Special Functions in Particle Physics}, Ph.D. Thesis, J. Kepler University 
Linz, 2012,
  arXiv:1305.0687 [math-ph];\\
  {\it A Computer Algebra Toolbox for Harmonic Sums Related to Particle Physics}, Diploma Thesis, J. Kepler University 
Linz, 2009,
  arXiv:1011.1176 [math-ph].
%
\bibitem{Ablinger:2011te}
  J.~Ablinger, J.~Bl\"umlein and C.~Schneider,
  J.\ Math.\ Phys.\  {\bf 52} (2011) 102301
  [arXiv:1105.6063 [math-ph]].
%
\bibitem{Ablinger:2013cf}
  J.~Ablinger, J.~Bl\"umlein and C.~Schneider,
  J.\ Math.\ Phys.\  {\bf 54} (2013) 082301
  [arXiv:1302.0378 [math-ph]].
%
\bibitem{EMSSP}
  J.~Ablinger, J.~Bl\"umlein, S.~Klein and C.~Schneider,
  Nucl.\ Phys.\ Proc.\ Suppl.\  {\bf 205-206} (2010) 110
  [arXiv:1006.4797 [math-ph]];\\
  J.~Bl\"umlein, A.~Hasselhuhn and C.~Schneider,
  PoS (RADCOR 2011) 032
  [arXiv:1202.4303 [math-ph]];\\
  C.~Schneider,
  J.\ Phys.\ Conf.\ Ser.\  {\bf 523} (2014) 012037
  [arXiv:1310.0160 [cs.SC]].
%
\bibitem{Vermaseren:1994je}
  J.A.M.~Vermaseren,
  Comput.\ Phys.\ Commun.\  {\bf 83} (1994) 45.
%
\bibitem{HSUM}
  J.A.M.~Vermaseren,
  Int.\ J.\ Mod.\ Phys.\ A {\bf 14} (1999) 2037
  [hep-ph/9806280]. \\
  J.~Bl\"umlein and S.~Kurth,
  Phys.\ Rev.\  D {\bf 60} (1999) 014018
  [arXiv:hep-ph/9810241].
%
\bibitem{Ablinger:2011pb}
  J.~Ablinger, J.~Bl\"umlein, S.~Klein, C.~Schneider and F.~Wi\ss{}brock,
  arXiv:1106.5937 [hep-ph].
%
\bibitem{Ablinger:2012qj}
  J.~Ablinger, J.~Bl\"umlein, A.~Hasselhuhn, S.~Klein, C.~Schneider and F.~Wi\ss{}brock,
  PoS {(RADCOR2011)} 031
  [arXiv:1202.2700 [hep-ph]].
%
\bibitem{Blumlein:2009rg}
  J.~Bl\"umlein, S.~Klein and B.~T\"odtli,
  Phys.\ Rev.\ D {\bf 80} (2009) 094010
  [arXiv:0909.1547 [hep-ph]].
%
\bibitem{MB}
  M.~Czakon,
  Comput.\ Phys.\ Commun.\  {\bf 175} (2006) 559
  [hep-ph/0511200];\\
%
\bibitem{MBr}
  A.V.~Smirnov and V.A.~Smirnov,
  Eur.\ Phys.\ J.\ C {\bf 62} (2009) 445
  [arXiv:0901.0386 [hep-ph]].
%
\bibitem{Bierenbaum:2007dm}
  I.~Bierenbaum, J.~Bl\"umlein and S.~Klein,
  Phys.\ Lett.\ B {\bf 648} (2007) 195
  [hep-ph/0702265].
%
\bibitem{Ablinger:2017bjx}
  J.~Ablinger, J.~Bl\"umlein, A.~De Freitas, M.~van Hoeij, E.~Imamoglu, C.~G.~Raab, C.-S.~Radu and C.~Schneider,
  arXiv:1706.01299 [hep-th].
%
\bibitem{Ablinger:2017cin}
  J.~Ablinger {\it et al.},
  PoS(RADCOR2017)069  arXiv:1711.09742 [hep-ph].
\end{thebibliography}
\end{document}